\documentclass[12pt,preprint]{aastex}
\usepackage{natbib}

\shorttitle{Nearby Southern Red Dwarfs}
\shortauthors{Winters et al.}

\begin{document}

\title{The Solar Neighborhood XXXV: Distances to 1404 M Dwarf Systems
  Within 25 pc in the Southern Sky}

\author{Jennifer G. Winters\altaffilmark{1}, Wei-Chun
  Jao\altaffilmark{1}, Sergio B. Dieterich\altaffilmark{1}}

\affil{Department of Physics and Astronomy, Georgia State University,
  Atlanta, GA 30302-4106}\email{winters@astro.gsu.edu,
  jao@astro.gsu.edu, dieterich@astro.gsu.edu}

\author{John C. Lurie\altaffilmark{1}} 

\affil{Astronomy Department, University of Washington, Seattle, WA
  98195}\email{lurie@uw.edu}

\author{Todd J. Henry\altaffilmark{1}, Altonio
  D. Hosey\altaffilmark{1}, Kenneth J. Slatten, Mark
  R. Boyd\altaffilmark{1}, Philip A. Ianna\altaffilmark{1}}

\affil{RECONS Institute, Chambersburg, Pennsylvania,
  17201}\email{thenry@astro.gsu.edu, hosey@astro.gsu.edu,
  kslatten@cpinternet.com, boyd@astro.gsu.edu, philianna3@gmail.com}

\author{Nigel C. Hambly}

\affil{Scottish Universities Physics Alliance (SUPA), Institute of
Astronomy, University of Edinburgh, Royal Observatory, Blackford Hill,
Edinburgh EH9 3HJ, Scotland, UK}\email{nch@roe.ac.uk}

\author{Adric R. Riedel\altaffilmark{1}}

\affil{Hunter College/American Museum of Natural History, NYC,
NY}\email{ar494@hunter.cuny.edu}

\author{John P. Subasavage\altaffilmark{1}}

\affil{US Naval Observatory, Flagstaff Station, 10391 West Observatory
Road, Flagstaff, AZ 86001}\email{jsubasavage@nofs.navy.mil}

\and

\author{Charlie T. Finch\altaffilmark{1}}

\author{Jennifer L. Bartlett\altaffilmark{1}}

\affil{US Naval Observatory, 3450 Massachussetts Ave. NW, Washington,
  DC 20392}\email{finch@usno.navy.mil,
  jennifer.bartlett@usno.navy.mil}

\altaffiltext{1}{Visiting Astronomer, Cerro Tololo Inter-American
Observatory.  CTIO is operated by AURA, Inc.\ under contract to the
National Science Foundation.}
 
\begin{abstract}
\label{abstract}

We present trigonometric, photometric, and photographic distances to
1748 southern ($\delta \leq$0$^\circ$) M dwarf systems with $\mu \ge$
0\farcs18 yr$^{-1}$, of which 1404 are believed to lie within 25
parsecs of the Sun.  The stars have 6.67 $\leq$ $V_J$ $\leq$ 21.38 and
3.50 $\leq$ ($V_J-K_s$) $\leq$ 9.27, covering the entire M dwarf
spectral sequence from M0.0V through M9.5V.  This sample therefore
provides a comprehensive snapshot of our current knowledge of the
southern sky for the nearest M dwarfs that dominate the stellar
population of the Galaxy.  Roughly one-third of the 1748 systems, each
of which has an M dwarf primary, have published high quality
parallaxes, including 179 from the RECONS astrometry program.  For the
remaining systems, we offer photometric distance estimates that have
well-calibrated errors.  The bulk of these ($\sim$700) are based on
new $V_JR_{KC}I_{KC}$ photometry acquired at the CTIO/SMARTS 0.9m
telescope, while the remaining 500 primaries have photographic plate
distance estimates calculated using SuperCOSMOS $B_JR_{59F}I_{IVN}$
photometry.  Confirmed and candidate subdwarfs in the sample have been
identified, and a census of companions is included.


\end{abstract} 

\keywords{stars: distances --- stars: low mass --- stars: statistics
  --- solar neighborhood --- techniques: photometric --- parallaxes}

\section{Introduction}
\label{sec:intro}

There is great interest in developing a comprehensive list of nearby M
dwarfs for studies of the Milky Way's stellar population, both as
astrophysically compelling individual objects and as prime targets for
exoplanet and SETI search lists.  These stars, typically called red
dwarfs, dominate the stellar population of our Galaxy, accounting for
more than 75\% of all main sequence stars in the solar neighborhood
\citep{Henry(2006)}.\footnote{Updated counts provided at {\it
    www.recons.org}.}  Yet, distance determinations for red dwarfs
have remained a challenge.  Trigonometric parallaxes are the optimal
tool for calculating distances, but because of observing time and
resource constraints, distance estimation methods are attractive
alternatives and are often a preliminary step in determining which
targets are to be added to parallax programs.

There are two primary sources of trigonometric parallax data currently
available.  The {\it Yale Trigonometric Parallaxes, Fourth Edition}
\citep{vanAltena(1995)}, often called the {Yale Parallax Catalogue}
(hereafter YPC), is a valuable compendium of ground-based parallaxes
published prior to 1995 and includes about half of the southern M
dwarf parallaxes measured to date.  The {\it Hipparcos} mission
(initial release by \citet{Perryman(1997)}, and the updated results by
\citet{vanLeeuwen(2007)}; hereafter HIP) updated many of those
parallaxes, and contributed almost two hundred new measurements for
mostly bright ($V$ $\lesssim$~12.5) southern M dwarfs.

Since then, the RECONS (REsearch Consortium On Nearby Stars) group has
measured a substantial number of parallaxes to nearby stars to date,
adding 154 M dwarf systems to the 25 pc census
\citep{Jao(2005),Costa(2005),Henry(2006), Costa(2006),
  Subasavage(2009), Riedel(2010), Jao(2011), Riedel(2011),
  vonBraun(2011), Mamajek(2013), Jao(2014), Dieterich(2014),
  Riedel(2014)}, published in {\it The Solar Neighborhood} series of
papers (hereafter TSN) in {\it The Astronomical Journal}.  This effort
has increased the number of M dwarf systems with accurate
trigonometric parallaxes in the southern sky by 35\%, with several
hundred more determinations underway.

Both in addition to and in combination with proper motion surveys,
optical photometry has historically been used to identify and
characterize nearby stars.  The first generation of sky surveys for
intrinsically faint, nearby stars included the work of
\citet{Luyten(1979a), Luyten(1979b), Luyten(1980a), Luyten(1980b)} and
\citet{Giclas(1971), Giclas(1978)}.  The work continued into the late
twentieth century via the efforts of \citet{Bessel(1990),
  Bessell(1991)}, \citet{Weis(1984), Weis(1986), Weis(1987),
  Weis(1988), Weis(1991b), Weis(1991a), Weis(1993), Weis(1994),
  Weis(1996), Weis(1999)}, \citet{Wroblewski(1989), Wroblewski(1990),
  Wroblewski(1991), Wroblewski(1992), Wroblewski(1994),
  Wroblewski(1995), Wroblewski(1996), Wroblewski(1997),
  Wroblewski(1998)}, and \citet{Wroblewski(1999), Wroblewski(2000),
  Wroblewski(2001)}.  More recent surveys include the SAAO group
\citep{Kilkenny(1995), Kilkenny(1998), Kilkenny(2007), Koen(2002),
  Koen(2010), Reid(2001), Reid(2002), Reid(2003), Reid(2004)},
L{\'e}pine \citep{Lepine(2002), Lepine(2003), Lepine(2005b),
  Lepine(2005a), Lepine(2008), Lepine(2011)}, the USNO
\citep{Finch(2010), Finch(2012)}, and Deacon \citep{Deacon(2005b),
  Deacon(2005a), Deacon(2007), Deacon(2009a), Deacon(2009b)}, as well
as RECONS' efforts in the southern sky \citep{Hambly(2004),
  Henry(2004), Subasavage(2005a), Subasavage(2005b), Finch(2007),
  Winters(2011), Boyd(2011b), Boyd(2011a)}. This work is a
continuation in that tradition, and includes a comprehensive
assessment of the known population of red dwarfs in the southern sky,
with a particular emphasis on those closer than a 25 pc horizon.

\section{Definition of the Sample}
\label{sec:sampledef}

We have developed a list of 1748 southern ($\delta \leq$0$^\circ$)
stellar systems containing M dwarf primaries.\footnote{We refer to any
  collection of stars and their companion brown dwarfs and/or
  exoplanets as a system, including single M dwarfs not currently
  known to have any companions. Systems that contain a white dwarf
  component have been omitted, as the white dwarf was initially the
  brighter primary component.}  Here we define an M dwarf to be a star
with 3.50 $\leq$ ($V_J-K_s$) $\leq$ 9.27, corresponding to spectral
types M0.0V to M9.5V, where the red cut-off has been defined by
\citet{Henry(2004)}.  This list is a combination of objects with
existing high quality parallaxes, objects for which we have measured
photometry as part of our southern astrometry/photometry program, and
known objects that were recovered during our SuperCOSMOS-RECONS (SCR)
proper motion searches for new nearby stars. To eliminate any red
giants that may slip into this sample and to be consistent with the
proper motion cut-off of Luyten, we limit our targets to those with
proper motions, $\mu$, greater than 0\farcs18 yr$^{-1}$.



Table 1 provides the observed data for the entire sample, including
the name of the M dwarf primary, the number of known components in the
system, coordinates (J2000.0), proper motion magnitude and position
angle with reference, the weighted mean of the published trigonometric
parallaxes and the error, the number of parallaxes included in the
weighted mean and references, SuperCOSMOS $B_JR_{59F}I_{IVN}$ plate
magnitudes (hereafter simply $BRI$), $V_JR_{KC}I_{KC}$ (hereafter
simply $VRI$) measured by RECONS at Cerro Tololo Inter-American
Observatory (CTIO) and the number of observations, or $VRI$ from a
trusted source and the reference, and 2MASS $JHK_s$ magnitudes.  The
table is divided into the 1404 systems within 25 pc (top) and the 344
systems beyond 25 pc (bottom). A number in parentheses next to the
number of components column indicates how many of the components are
included in the photometry. All proper motions are from SuperCOSMOS,
except where noted.\footnote{Some of the proper motion values from the
  SuperCOSMOS Sky Survey for SCR objects presented here differ
  slightly from those in the original discovery papers.  For the new
  values, a more comprehensive method has been used that provides
  multiple measurements of proper motion between various plate pairs,
  while before only a single value was available that incorporated all
  plates into the solution simultaneously.  The proper motions
  provided here are those that yield consistent values among multiple
  determinations and are preferred.} '$J$' next to a magnitude
indicates that light from a close companion has resulted in blended
photometry.  A 'u' next to the photometry reference code indicates
that we have previously published photometry for this object, but have
since then acquired more data and updated the value.

Each system has at least one of three different distance
determinations, listed in Table 2, with the best distance either
photographic ({\it pltdist}, 500 systems), photometric ({\it ccddist},
667), or trigonometric ({\it trgdist}, 581).\footnote{The term {\it
    pltdist} is used for the remainder of the paper to indicate a
  distance estimate that combines SuperCOSMOS plate
  $B_JR_{59F}I_{IVN}$ and 2MASS $JHK_s$ magnitudes, as discussed in
  $\S$ 4.1. The plate $R_{59F}$ magnitude is the second, more recent
  $R$ epoch measurement of the two available in the SuperCOSMOS Sky
  Survey. The term {\it ccddist} is used to indicate a distance
  estimate that combines CCD $V_JR_{KC}I_{KC}$ and 2MASS $JHK_s$
  magnitudes, as discussed in $\S$4.2. The term {\it trgdist} is used
  for those objects for which an accurate trigonometric parallax has
  been measured, as discussed in $\S$ 3.4.}  Most of the systems
presented here have more than one type of distance measurement.  For
example, a system with a trigonometric parallax published by our group
is likely to also have a {\it ccddist} and a {\it pltdist} because we
first estimated a distance photometrically from SuperCOSMOS plates,
then measured $VRI$ photometry to yield a more accurate {\it ccddist},
then observed the system for a parallax measurement, which provides
{\it trgdist}.  We rank the quality of the distances by the errors
associated with each method: {\it pltdists} from photographic
photometry rank third (errors at least 26\%), {\it ccddists} from
$VRI$ photometry rank second (errors at least 15\%), and {\it
  trgdists} derived from accurate trigonometric parallaxes are the
best available (errors $<$5\%). Figure 1 illustrates the distribution
of all 1748 systems on the sky.


\subsection{Multiples}
\label{sec:disc.multiples}

We are currently involved in an all-sky study of the multiplicity of M
dwarfs using a sample of $\sim$1300 red dwarfs having accurate
trigonometric parallaxes placing them within 25 pc. This project
involves three search methods to identify and confirm companions at
separations from 0.1$\arcsec$ to 600$\arcsec$: a robotic adaptive
optics search (0.1--2$\arcsec$), an $I$-band imaging search
($\sim$1--10\arcsec), and a common proper motion search
($\sim$5--600\arcsec). Because most of these results will be published
in a future paper, we give only the number of known components in each
system in Table 1. An exhaustive literature search was not performed.
Based on this cursory search, the number of known multiples in the 25
pc portion of the sample is 164, which results in a multiplicity
fraction of 12\%.  This is one-half to one-third the value that is
expected \citep{Henry(1991), Fischer(1992), Janson(2012),
  Janson(2014)}, so it is clear that more work is needed to determine
the true multiplicity fraction of even the nearest M dwarfs.  Of the
164 multiples, 114 have {\it pltdists} that are based on unresolved
photometry.  These are typically multiples with magnitude differences
at $BRI$ $\lesssim$~2 and with separations $<$4$\arcsec$.  Of these 114
with {\it pltdists}, 96 systems have {\it ccddists} based on
unresolved photometry because their separations are $<$1$\arcsec$.  In
addition, photometry from the SAAO extracted from the literature
(references in $\S$1) uses apertures 21--31$\arcsec$ in diameter, so
additional components have likely been included in the photometric
values used to derive {\it ccddists}.  The comprehensive survey of red
dwarf multiplicity mentioned above constitutes a major fraction of the
first author's Ph.D. work, so a much more careful treatment of red
dwarf multiples within 25 pc is forthcoming. For now, we note that the
reader should be cautious when applying the {\it pltdist} or {\it
  ccddist} values given in Table 2 for multiple systems.

\section{Data}
\label{sec:data}
\subsection{$B_JR_{59F}I_{IVN}$ Plate Photometry}
\label{sec:plate.phot}

Plate magnitudes from SuperCOSMOS are given in Table 1 for all but a
few systems and are rounded to the nearest hundredth
magnitude.\footnote{The wavelength ranges for the $B_J$, R$_{59F}$,
  and I$_{IVN}$ filters are 3950-5400\AA, 5900-6900\AA, and
  7150-9000\AA, respectively \citep{Morgan(1995)}.}  Errors are
typically 0.3 mags for magnitudes fainter than 15, with larger errors
for brighter objects due to systematic errors
\citep{Hambly(2001)}. Derived {\it pltdists} in Table 2 for SCR
discoveries have been previously presented in the TSN papers, but we
provide 1457 new {\it pltdists} here for previously known objects.
Additional SuperCOSMOS queries were recently done in an effort to
provide as many {\it pltdists} as possible. Due to an improved
SuperCOSMOS photometric calibration,\footnote{The recommended access
  point for the current version of the photometrically calibrated
  SuperCOSMOS data is the SuperCOSMOS Science Archive, found at {\it
    http://surveys.roe.ac.uk/ssa.} Details of the photometric
  calibration procedure are available in the online documentation at
  {\it http://surveys.roe.ac.uk/ssa/dboverview.html\#mags}.} some
magnitudes will be slightly different from those reported in previous
papers in this series; the values presented here are preferred. In
some cases, a match was not found because very red objects were not
always recovered from the $B$ plate, because of very high proper
motion, or because of source mergers or corruption.  All matches were
visually confirmed and were doublechecked using the SuperCOSMOS proper
motions and $BRI$ magnitudes.

\subsection{$V_JR_{KC}I_{KC}$ CCD Photometry}
\label{sec:ccd.phot}

We have measured $VRI$ photometry for 799 of the systems presented
here as a part of our astrometry/photometry program, primarily at the
CTIO/SMARTS 0.9m, with a few measurements from the CTIO/SMARTS 1.0m
(see \citet{Jao(2005), Henry(2006)} for details on the
astrometry; see \citet{Winters(2011)} for details on the photometry).
The observations were made between 1999 and 2013.  The photometry was
typically acquired for these nearby star candidates based upon
{\it pltdists} from our SuperCOSMOS trawls or from other proper motion or
photometric surveys such as those listed in $\S$1.

All of our $VRI$ data, given in Table 1, were reduced using IRAF and
are on the Johnson-Kron-Cousins system.\footnote{The central
  wavelengths for the $V_J$(old), V$_J$(new), R$_{KC}$, and I$_{KC}$
  filters are 5438\AA, 5475\AA, 6425\AA, and 8075\AA, respectively.
  See \citet{Jao(2011)} for a discussion of the nearly identical
  $V_J$(old) and V$_J$(new) filters.}  Calibration frames taken at the
beginnings of nights were used for typical bias subtraction and dome
flat-fielding.  Standard star fields from \citet{Graham(1982)},
\citet{Bessel(1990)}, and/or \citet{Landolt(1992), Landolt(2007),
  Landolt(2013)} were observed multiple times each night in order to
derive transformation equations and extinction curves.  In order to
match those used by Landolt, apertures 14$\arcsec$~in diameter were
used to determine the stellar fluxes, except in cases where close
contaminating sources needed to be deblended.  In these cases, smaller
apertures were used and aperture corrections were applied.  As
outlined in \citet{Winters(2011)}, photometric errors are typically 30
millimagnitudes (mmags) for the $V$-band and 20 mmags for both the
$R$- and $I$-bands.  Further details about the data reduction
procedures, transformation equations, etc., can be found in
\citet{Jao(2005)} and \citet{Winters(2011)}.

An additional 369 primaries were found to have high quality optical
photometry available from the literature, primarily from Bessell, Weis,
and the SAAO group.\footnote{References given in Table 1.} Among
these, 214 already have parallaxes that place them within 25 pc, while
an additional 109 have a {\it ccddist} that place them within 25 pc.
The $R$ and $I$ photometric values from Weis have all been transformed
to the Johnson-Kron-Cousins system via the color relations given in
\citet{Bessell(1987)} so that all $VRI$ values in Table 1 are on the
same photometric system. SOAR/SOI (SOAR Optical Imager) photometry
from \citet{Dieterich(2014)} has also been converted to the
Johnson-Kron-Cousin system using methods described in that paper.


A comparison of ($V_{CCD}-K_s$) versus ($B_{plt}-K_s$) for 1026 objects
for which both plate $B$ and CCD $V$ photometry is available is shown
in Figure 2. Known close binaries for which the photometry is blended
have been omitted. A third order polynomial fit,

($V_{J}-K_s$) $=$ 4.939 $-$ 1.174*($B_{plt}-K_s$) $+$ 0.256*($B_{plt}-K_s$)$^2$ $-$ 0.010*($B_{plt}-K_s$)$^3$ 

\hskip-25pt permits users of SuperCOSMOS data to predict a CCD $V_{J}$
magnitude from a given ($B_{plt}-K_s$) for red objects similar to
those given in this paper, ($V_J-K_s$) $=$ 3.5--9.27 or ($B_{plt}-K_s$)
$\approx$ 3.7--10.7, assuming a 2MASS $K_s$ magnitude is known.

\subsection{$JHK_s$ Photometry from 2MASS}
\label{sec:IR.phot}

Infrared photometry in the $JHK_s$ (hereafter simply $JHK$) system has
been extracted from 2MASS \citep{Skrutskie(2006)} and is rounded to
the nearest hundredth magnitude in Table 1.  All magnitudes have been
checked by eye. As described in $\S$4, the same 2MASS photometry has
been used for both the {\it pltdists} and the new {\it ccddists}
presented here. Errors are typically less than 50 mmags.  Exceptions
are indicated in Table 1 as superscripts to the magnitudes.

\subsection{Trigonometric Parallaxes}
\label{sec:trigdist}

A total of 442 southern M dwarf systems have {\it trigdists} placing
them within 25 pc; their weighted mean parallaxes and resulting
distances are listed in Tables 1 and 2.  These systems have been
extracted from the new RECONS 25 Parsec Database that contains all
stellar systems with published trigonometric parallaxes of at least 40
mas and with errors less than 10 mas.  The RECONS group has published
the only parallax for 147 (25$\%$) of the 581 systems with parallaxes
reported in this paper. Included in that 147 are 24 systems in the
lower portions of Tables 1 and 2 that have parallaxes $<$ 40 mas
measured by RECONS --- these were anticipated to be closer than 25 pc
because of their {\it pltdists} and/or {\it ccddists}, but are now
known to lie beyond 25 pc.


For all systems without parallaxes, a search of YPC and HIP was
carried out in the event that a parallax $<$ 40 mas already existed.
A 30$\arcmin$ search radius for YPC and a 5$\arcmin$ search radius for
HIP were used in sweeps for objects to compensate for high proper
motions as well as poor coordinates, the latter being particularly
important for the YPC.  Possible matches were then confirmed or
refuted by comparing identifiers, proper motions, and $V$ magnitudes.
Seven percent (114 systems) of the entire sample in this paper were
discovered to have parallaxes $<$ 40 mas in YPC and/or HIP.  No other
systematic literature search for stars with published parallaxes
beyond 25 pc has yet been done beyond those using YPC and HIP.

\section{Distances}
\label{sec:distances}
\subsection{Photographic Plate Distance Estimates}
\label{sec:platedist}

The {\it pltdist} estimates in Table 2 are calculated by combining
SuperCOSMOS $BRI$ photometry with 2MASS infrared photometry via a
suite of 15 color-$M_{K}$ relations using $BRIJHK$, as described in
\citet{Hambly(2004)}.  The four relations $M_K$ vs.~$B-R$, $J-H$,
$J-K$, and $H-K$ have limited spans in color through the M dwarf
sequence and are thus omitted.  The {\it pltdist} estimate is
considered reliable if the remaining 11 relations are applicable,
i.e.~if a star's color falls within the range covered by the
calibrations for single, main sequence stars.  However, if a target
star is blended with another source on one plate, up to five relations
may drop out of the suite, yielding a less reliable {\it pltdist}
based on 6--10 relations.  We consider six relations to be the minimum
number acceptable for a {\it pltdist} because at least two of the
three $BRI$ magnitudes, combined with three 2MASS measurements,
provide optical/infrared colors consistent with those of normal main
sequence stars.  A few stars in Table 2 have {\it pltdists} with fewer
than six relations because plate magnitudes were extracted and
distances estimated after it was known that $VRI$ photometry and/or
trigonometric parallaxes were available. Thus the new extractions of
SuperCOSMOS data simply augment the sample and provide as many
distances as possible for comparisons.

As described in \citet{Hambly(2004)}, to estimate the reliability of
the {\it pltdists} generated from the suite of relations, single, main
sequence, M dwarfs with known {\it trgdists} were run back through the
relations to derive representative errors.  The mean offsets between
the {\it pltdists} and {\it trigdists} were found to be 26\%.  In
Table 2 we list the total errors that include this systematic value
combined in quadrature with the standard deviation of the up to 11
individual distances computed for a given star.

\subsection{CCD Distance Estimates}
\label{sec:ccddist}

The {\it ccddists} in Table 2 are determined using a method similar to
that used for the {\it pltdists} and are described in
\citet{Henry(2004)}.  The difference is that we use more accurate CCD
$VRI$ magnitudes obtained at CTIO instead of plate magnitudes from
SuperCOSMOS to determine the suite of color-$M_K$ relations.  Again,
the maximum number of possible relations from the combination of
$VRIJHK$ magnitudes is 15, but in this case 12 relations yield useful
results, as the color spreads in $M_K$ vs.~$J-H$, $J-K$, and $H-K$ are
limited and these relations are omitted from the suite.  For stars
with $VRI$ from the literature rather than from our observing program,
the {\it ccddists} were calculated using the same relation suite.  All
photometry is on the Johnson-Kron-Cousins system, and therefore the
resulting {\it ccddists} are all generated in a uniform fashion.

Similar to the {\it pltdist}, stars with all 12 relations have {\it
  ccddists} we deem reliable (assuming the stars are single and on the
main sequence), and those with 7--11 relations we deem suspect because
at least one magnitude and up to five relations have dropped out of
the suite.  Again, some stars in Table 2 have {\it ccddists} with
fewer than six relations because $VRI$ photometry was gathered and
distances estimated after it was known that trigonometric parallaxes
were available. Thus the new extractions of $VRI$ data simply augment
the sample and provide as many distances as possible for comparisons.
As described in \citet{Henry(2004)}, to estimate the reliability of
the {\it ccddists} generated from the suite of relations, single, main
sequence M dwarfs with known {\it trgdists} were run back through the
relations to derive representative errors.  The mean offsets between
the {\it ccddists} and {\it trigdists} were found to be 15\%.  In
Table 2 we list the total errors that include this systematic value
combined in quadrature with the standard deviation of the up to 12
individual distances computed for a given star.  Those objects with
equal magnitude companions included in unresolved photometry will have
distance estimates placing them too close by a factor of 1.4. Light
from fainter companions will decrease this offset, whereas light from
additional unresolved companions will increase this offset.


Table 2 provides the derived data, split into the 25 pc sample (top)
and systems with best quality distances beyond 25 pc (bottom).  Data
listed include the name of the M dwarf primary, coordinates,
information on the {\it pltdists} (the distance, the total error,
number of relations), {\it ccddists} (the distance, the total error,
number of relations), and {\it trgdists} (the distance from the
weighted mean of published parallaxes, the weighted mean error, and
the number of parallaxes used to generate the weighted mean).  These
empirical values are then followed by the most reliable distance and
its type.  Distances based upon blended photometry are given in square
brackets. Distances for subdwarfs (both confirmed and candidates) are
given in curly brackets and are typically closer than estimated.

\subsection{Distance Comparisons}
\label{sec:comparison}

Figure 3 shows a comparison of the {\it pltdists} and {\it ccddists}
for the 739 stars that have both $BRI$ and $VRI$ photometry placing
them within 25 pc.  Only those objects thought to be single have been
plotted. Error bars on individual points are omitted for clarity, but
can be found for individual systems in Table 2.  Total errors are
typically 32\% in {\it pltdist} and 16\% in {\it ccddist}, and include
the computed systematic errors inherent to the two techniques
(26\% and 15\% for {\it pltdists} and {\it ccddists}, respectively),
and the standard deviations of the individual estimates for each star.
Because estimates from individual relations in the suites are usually
quite consistent, the total errors are dominated by the adopted values
for systematic errors.  The dominant cause of the distance
discrepancies is poor plate photometry compared to CCD photometry.


Figures 4 (343 systems) and 5 (337 systems) compare the {\it pltdists}
and {\it ccddists}, respectively, to the available {\it trgdists}
placing systems within 25 pc, again using only single objects.  The
average offsets are 24\% for the {\it pltdists} and 17\% for the {\it
  ccddists},consistent with the 26\% and 15\% reported in
\citet{Hambly(2004)} and \citet{Henry(2004)}. The inherent spreads
around the one-to-one lines in each plot have three causes: (1) errors
in the photometry and parallaxes used to derive the relations and for
the stars targeted here, (2) unresolved systems, and (3) cosmic
scatter due to differences in stellar ages, compositions, and perhaps
magnetic properties among the stars that are not taken into account by
the color-magnitude relations.  $BRI$ photometry errors dominate the
{\it pltdist} vs.~{\it trgdist} offsets, while cosmic scatter
dominates the {\it ccddist} vs.~{\it trgdist} offsets because the
$VRI$ photometry and parallax errors are each only a few percent.  The
reduced scatter in Figure 5 again illustrates the value of obtaining
accurate photometry, as the scatter has been reduced by roughly a
factor of two. While it appears that the distances are more often
underestimated than overestimated, this is a result of the presence of
unresolved multiples and young objects in the sample.


\section{Results}
\label{sec:results}
\subsection{M Dwarfs, M Subdwarfs and M Giants}
\label{sec:colors}

Before outlining the population results of this study, we must first
verify that stars included in the final sample are, indeed, M dwarfs
based on the photometry and parallaxes now available.  Although our
sample has been selected to include only stars with proper motions in
excess of $\mu$ = 0\farcs18 yr$^{-1}$, a few giants might be included
that have extraordinary space velocities, or more likely, have
erroneous proper motions.  For stars without parallaxes, we use
several color-color diagrams to separate dwarfs from interloping
giants --- an example is shown in Figure 6, which plots ($J-H$)
vs.~($R-K$).  There are some objects in this sample that have a
published trigonometric parallax, but for which reliable CCD $VRI$
photometry does not yet exist. Those objects have been included on
this plot using the plate $R$ magnitude, rather than on the
absolute-color diagram in Figure 7.

A small supplementary set of giants not in the M dwarf sample
discussed here is included for comparison.  These giants have been
observed at the 0.9m using the same instrument configuration as for
the dwarfs. Only those objects with 2MASS magnitude quality codes of
{\it AAA} are used in Figure 6. While there appears to be some
blending of giants and dwarfs in the early M section of the plot
($(R-K)$ $\sim$ 2.5-3.5 and $(J-H)$ $\sim$0.7-0.8), it is unlikely
that any of the presumed dwarfs with only a photometric distance
estimate in that region will be giants, given the proper motion limit
of this sample. It is notable that the giants separate very cleanly
from the dwarfs redward of $(R-K)$ $\sim$4. The $VRIJHK$ photometry
for these objects is listed in Table 3.

Four stars in this diagram are worthy of note. We have been following
the initially very interesting target CD-32 16735 on our parallax
program and have measured for it a preliminary parallax of $-$4.69
$\pm$ 3.88 mas, essentially zero within the errors. Thus, this object
is most certainly a giant. Both L173-003 and GJ0552.1 are likely
unresolved binaries with odd colors from an inspection of their
SuperCOSMOS images (as equal luminosity binaries will appear normal on
a color-color plot). A comparison of the two available distances for
L173-003 also indicates that it is an unresolved binary, as its {\it
  pltdist} (23.4 pc) is an underestimate of the true {\it trgdist}
(47.9 pc). GJ0552.1 also has two distances available for comparison;
however they are both photometric estimates that agree within the
errors: {\it pltdist} $=$ 13.2 $\pm$ 5.3 versus {\it ccddist} $=$ 16.3
$\pm$ 3.9. HD320012 has only a {\it pltdist}, but has large $H$ and
$K$ magnitude errors.


For the roughly one-third of systems with parallaxes, confirming that
they are red dwarfs is straightforward using an empirical HR diagram.
An example, using $M_{V}$ vs.~($V-K$), is shown in Figure 7.  The M
dwarf sequence is well-defined, and there is a noticeable population
of cool subdwarfs below the main sequence having ($V-K$) = 3.5--4.5.
As is evident from this diagram, none of the stars with trigonometric
parallaxes presented in this paper fall in the region where known
giants are found, so we are confident that at least among the stars
with parallaxes, there are no giants.


In Figure 8 we show a reduced proper motion diagram
\citep{Boyd(2011a)} to confirm the subdwarfs identified in Figure 7
and to find additional subdwarf candidates in the sample that do not
have parallaxes.  Note that because subdwarfs are found below the main
sequence (contrary to the positions of young and multiple objects
above the main sequence), their distances will be {\it overestimated}
rather than underestimated.  Distances for these objects have been
enclosed in curly brackets in Table 2 to highlight this effect.  In
total, we have identified 24 red subdwarf candidate systems, listed in
Table 4.  Of these 24 candidates, 21 have parallaxes, three do not,
and 18 are confirmed subdwarfs (LHS2852 is buried in the mass of
clustered points). While LHS3528, LHS0284, and LHS0323 both appear in
(or on the border of) the subdwarf region of Figure 8, they have been
shown spectroscopically to be dwarfs
\citep{Hawley(1996),Jao(2011)}. LHS0110 and SSS1444-2019 have both
been identified as candidates by others \citep{Jao(2011),
  Schilbach(2009)}.

%

\subsection{The Red Dwarf Population}
\label{sec:population}

Of the 1748 systems presented here, our best distance estimates
place 1404 closer than 25 pc.  Table 5 lists the distance
statistics for the entire sample.  Those primaries with distance
underestimates due to blended photometry are given in brackets next to
the numbers in the multiples column.  The 344 systems beyond 25 pc
were typically pushed over the horizon via new $VRI$ and/or new
parallax measurements, or are supplementary entries to the sample with
new $VRI$ photometry.  Even if an updated and better quality distance
has moved a system beyond 25 pc, it remains in the sample for distance
comparison purposes.

This sample is of sufficient size to calculate ``retention rates'' for
stars that were first estimated to be within defined distance limits
via the two photometric distance estimating techniques.


For the 1512 primaries that have {\it pltdists} that place them within
25 pc, 417 now have a trigonometric parallax ($\pi_{trig}$) $\geq$~40
milliarcseconds (mas), and 112 now have $\pi_{trig}$ $<$~40 mas,
resulting in a 78.8\% likelihood that an object initially estimated to
be within 25 pc will be found within 25 pc. For the 904 primaries with
{\it pltdists} $\leq$ 20 pc, 379 now have $\pi_{trig}$ $\geq$~40 mas,
and 67 now have $\pi_{trig}$ $<$~40 mas, leading to a retention rate
of 85.0\%. Finally, for the 443 systems with {\it pltdists} $\leq$ 15
pc, 276 now have $\pi_{trig}$ $\geq$~40 mas, and 18 now have
$\pi_{trig}$ $<$~40 mas, giving a retention rate of 93.9\%.


For the 896 primaries that have {\it ccddists} that place them within
25 pc, 402 now have $\pi_{trig}$ $\geq$~40 mas, and 32 now have
$\pi_{trig}$ $<$~40 mas, giving a retention rate of 92.6\%. For the
664 primaries with {\it ccddists} $\leq$ 20 pc, 359 now have
$\pi_{trig}$ $\geq$~40 mas, and 12 now have $\pi_{trig}$ $<$~40 mas,
leading to a retention rate of 96.8\%. And for the 409 systems with
{\it ccddists} $\leq$ 15 pc, 257 now have $\pi_{trig}$ $\geq$~40 mas,
and 4 now have $\pi_{trig}$ $<$~40 mas, resulting in a 98.5\%
likelihood that an object initially estimated to be within 15 pc will
be found within 25 pc.

Thus, our photometric distance estimating techniques are successful in
revealing red dwarfs within 25 pc 79--99\% of the time, depending on
which set of photometry is used and how restrictive the boundary is
set. Even in the worst case, we are successful in nearly 4 out of 5
cases.


Figure 9 outlines the volume density of the southern M dwarfs within
25 pc, using the ($V-K$) color versus distance in ten equal volume
shells to 25 pc.\footnote{For those $\sim$500 targets that lack a
  $V_J$ magnitude, one has been estimated using the polynomial given
  in $\S3.2$.}  We expect a uniform volume density to the 25 pc
horizon, and that is now the case for red dwarfs with types earlier
than $\sim$M4.  Although recent work by RECONS \citep{Dieterich(2014)}
has bolstered the number of the closest and reddest objects on this
plot, the sample shows a classic observational bias for redder,
i.e.~intrinsically fainter stars that are underrepresented at larger
distances.  There are several reasons for the non-uniform
distribution, including: (1) The reddest dwarfs with types later than
M6 have simply been missed in various search efforts, including our
SCR trawls, due to their intrinsic faintness.  (2) Some of the
currently known 25 pc members are hidden multiples --- resolution of
these multiples would move the points to larger distances if they have
been placed in Figure 9 using {\it pltdist} or {\it ccddist}. (3) Some
systems are likely among the $\sim$80 southern red dwarfs within 25
pc predicted to have $\mu$ $<$ 0\farcs18 yr$^{-1}$ by
\citet{Riedel(2012)}.  These have been excluded due to our proper
motion cut-off.  In fact, we currently have $\sim$40 stars on our
parallax program with $\mu$ $<$ 0\farcs18 yr$^{-1}$ having preliminary
parallaxes larger than 40 mas. (4) Finally, some red dwarfs are
lurking in the Galactic plane, which has traditionally been avoided
due to crowded fields, although this deficit is not likely to be
significantly distance-dependent.


We can assess the incompleteness of the current sample by assuming
that the 19 southern systems with M dwarf primaries found within 5 pc
\citep{Henry(2014)} represent a complete sample and that the stellar
density is constant to 25 pc.  If so, we expect 2375 such systems
within 25 pc in the southern sky.  The histogram in Figure 10 plots
the cumulative number of expected systems at distances to 25 pc as a
solid line, with the dotted lines representing the Poisson
errors. Here we have propagated the 23\% error on the 19 systems
within 5 pc to 25 pc, corresponding to an error of 546 systems in
2375.  The histogram of systems outlined for the trigonometric sample
is the starting point for systems we consider reliably within 25 pc.
The histograms for the CCD and plate additions include only those
systems with best distances that are ``clean'', i.e.~those that are
for single stars based on current information, and that are not
subdwarfs. These histograms are presumably overestimates, as a
fraction of the included systems are as yet undiscovered close
multiples with blended photometry whose plate and CCD distances are
underestimates.  We do not expect these overestimates to be extreme,
in part because some of the known close multiple systems purposely
removed to clean the sample will remain within 25 pc.  With these {\it
  caveats} in mind, we make the following assessments of our current
knowledge:

{\it We predict that there are 1829--2921 stellar systems with red
  dwarf primaries within 25 pc in the southern sky.}

{\it We have identified roughly 1400 of these systems.}

{\it Based on our retention rates, only $\sim$90\% of those with
  distance estimates or 1302 of these will remain within 25 pc.}

{\it There are $\sim$530--1620 systems missing from the current
  sample.}

Using the mass-M$_K$ relation given in \citet{Henry(1993)}, masses
have been estimated for each M dwarf primary in the sample that is
currently believed to be a single, main sequence star. The resulting
mass function, the number of stars within each mass bin, for southern
primaries within 25 pc is illustrated in Figure 11.  The mass function
may turn over at some point, but the turnover at $\sim$0.15M$_{\odot}$
indicated in this histogram is likely not the final answer. Early work
by \citet{Henry(1990)} pointed to $\sim$0.1 M$_{\odot}$ as the end of
the main sequence and where the mass function will turn over. More
recent work by \citet{Chabrier(2003)} and \citet{Dieterich(2012)}
reinforces this result. As is evident in Figure 9, few of the
intrinsically faintest M dwarfs have yet to be identified beyond
$\sim$20 pc, corresponding to roughly half of the sample volume, and
these are the stars that will fill the lowest mass bins in Figure 11.
Perhaps even more important, the true mass function for low mass stars
will include M dwarfs as companions, and both known and as yet unknown
fainter companions to M dwarf primaries in the sample will shift
points to lower masses.  This makes a profound statement that the
smallest stars are the most likely product of the stellar/substellar
formation process.


\section{Future}
\label{sec:future}

Of the 1404 southern M dwarf systems presented here as within 25 pc,
954 do not have published parallaxes.  More than 250 objects with only
distances estimates are on our astrometry program at the 0.9m.  If
these systems lacking a parallax turn out to be within 25 pc, the
RECONS effort in the southern sky will have increased the number of
stellar systems identified within this distance by 52\%.

Of course, {\it Gaia} should make great contributions to the census of
nearby red dwarfs.  However, assuming its $V$~$\approx$~20 limit
\citep{Sozzetti(2014)} is achieved, it will only reach the
intrinsically faintest M dwarfs to about 10 pc.  Thus, it appears that
careful, pointed observations to reveal the M dwarfs within 25 pc
remain warranted and will be crucial in determining the true mass
function for the smallest stars.

\section{Acknowledgments}

This research was made possible by NSF grants AST-0908402 and
AST-1109445.  The support of additional RECONS team members was vital
for this research.  We also thank the members of the SMARTS
Consortium, who enable the operations of the small telescopes at CTIO,
as well as the observer support at CTIO, specifically Edgardo
Cosgrove, Arturo Gomez, Alberto Miranda, Joselino Vasquez, Mauricio
Rojas, and Hernan Tirado. We would like to thank Rick Mascall for
bringing to our attention the object OUT1720-1725. The authors are
grateful to referee Bill van Altena for suggesting changes that
improved the paper.

This research has made use of data obtained from the SuperCOSMOS
Science Archive, prepared and hosted by the Wide Field Astronomy Unit,
Institute for Astronomy, University of Edinburgh, which is funded by
the UK Science and Technology Facilities Council. Data products from
the Two Micron All Sky Survey, which is a joint project of the
University of Massachusetts and the Infrared Processing and Analysis
Center/California Institute of Technology, funded by the National
Aeronautics and Space Administration and the National Science
Foundation have also been used, as have the SIMBAD database and the
Aladin and Vizier interfaces, operated at CDS, Strasbourg, France.


\bibliographystyle{apj}
\bibliography{msouth.references}

\begin{thebibliography}{130}
\expandafter\ifx\csname natexlab\endcsname\relax\def\natexlab#1{#1}\fi

\bibitem[{{Andrei} {et~al.}(2011){Andrei}, {Smart}, {Penna}, {d'Avila},
  {Bucciarelli}, {Camargo}, {Crosta}, {Dapr{\`a}}, {Goldman}, {Jones},
  {Lattanzi}, {Nicastro}, {Pinfield}, {da Silva Neto}, \&
  {Teixeira}}]{Andrei(2011)}
{Andrei}, A.~H., {et~al.} 2011, \aj, 141, 54

\bibitem[{{Anglada-Escud{\'e}} {et~al.}(2012){Anglada-Escud{\'e}}, {Boss},
  {Weinberger}, {Thompson}, {Butler}, {Vogt}, \&
  {Rivera}}]{Anglada-Escude(2012)}
{Anglada-Escud{\'e}}, G., {Boss}, A.~P., {Weinberger}, A.~J., {Thompson},
  I.~B., {Butler}, R.~P., {Vogt}, S.~S., \& {Rivera}, E.~J. 2012, \apj, 746, 37

\bibitem[{{Benedict} {et~al.}(2002){Benedict}, {McArthur}, {Forveille},
  {Delfosse}, {Nelan}, {Butler}, {Spiesman}, {Marcy}, {Goldman}, {Perrier},
  {Jefferys}, \& {Mayor}}]{Benedict(2002)}
{Benedict}, G.~F., {et~al.} 2002, \apjl, 581, L115

\bibitem[{{Bessel}(1990)}]{Bessel(1990)}
{Bessel}, M.~S. 1990, \aaps, 83, 357

\bibitem[{{Bessell}(1991)}]{Bessell(1991)}
{Bessell}, M.~S. 1991, \aj, 101, 662

\bibitem[{{Bessell} \& {Weis}(1987)}]{Bessell(1987)}
{Bessell}, M.~S., \& {Weis}, E.~W. 1987, \pasp, 99, 642

\bibitem[{{Biller} \& {Close}(2007)}]{Biller(2007)}
{Biller}, B.~A., \& {Close}, L.~M. 2007, \apjl, 669, L41

\bibitem[{{Boyd} {et~al.}(2011{\natexlab{a}}){Boyd}, {Henry}, {Jao},
  {Subasavage}, \& {Hambly}}]{Boyd(2011b)}
{Boyd}, M.~R., {Henry}, T.~J., {Jao}, W.-C., {Subasavage}, J.~P., \& {Hambly},
  N.~C. 2011{\natexlab{a}}, \aj, 142, 92

\bibitem[{{Boyd} {et~al.}(2011{\natexlab{b}}){Boyd}, {Winters}, {Henry}, {Jao},
  {Finch}, {Subasavage}, \& {Hambly}}]{Boyd(2011a)}
{Boyd}, M.~R., {Winters}, J.~G., {Henry}, T.~J., {Jao}, W.-C., {Finch}, C.~T.,
  {Subasavage}, J.~P., \& {Hambly}, N.~C. 2011{\natexlab{b}}, \aj, 142, 10

\bibitem[{{Chabrier}(2003)}]{Chabrier(2003)}
{Chabrier}, G. 2003, \pasp, 115, 763

\bibitem[{{Costa} \& {M{\'e}ndez}(2003)}]{Costa(2003)}
{Costa}, E., \& {M{\'e}ndez}, R.~A. 2003, \aap, 402, 541

\bibitem[{{Costa} {et~al.}(2005){Costa}, {M{\'e}ndez}, {Jao}, {Henry},
  {Subasavage}, {Brown}, {Ianna}, \& {Bartlett}}]{Costa(2005)}
{Costa}, E., {M{\'e}ndez}, R.~A., {Jao}, W.-C., {Henry}, T.~J., {Subasavage},
  J.~P., {Brown}, M.~A., {Ianna}, P.~A., \& {Bartlett}, J. 2005, \aj, 130, 337

\bibitem[{{Costa} {et~al.}(2006){Costa}, {M{\'e}ndez}, {Jao}, {Henry},
  {Subasavage}, \& {Ianna}}]{Costa(2006)}
{Costa}, E., {M{\'e}ndez}, R.~A., {Jao}, W.-C., {Henry}, T.~J., {Subasavage},
  J.~P., \& {Ianna}, P.~A. 2006, \aj, 132, 1234

\bibitem[{{Deacon} \& {Hambly}(2001)}]{Deacon(2001)}
{Deacon}, N.~R., \& {Hambly}, N.~C. 2001, \aap, 380, 148

\bibitem[{{Deacon} \& {Hambly}(2007)}]{Deacon(2007)}
---. 2007, \aap, 468, 163

\bibitem[{{Deacon} {et~al.}(2005{\natexlab{a}}){Deacon}, {Hambly}, \&
  {Cooke}}]{Deacon(2005b)}
{Deacon}, N.~R., {Hambly}, N.~C., \& {Cooke}, J.~A. 2005{\natexlab{a}}, \aap,
  435, 363

\bibitem[{{Deacon} {et~al.}(2005{\natexlab{b}}){Deacon}, {Hambly}, {Henry},
  {Subasavage}, {Brown}, \& {Jao}}]{Deacon(2005a)}
{Deacon}, N.~R., {Hambly}, N.~C., {Henry}, T.~J., {Subasavage}, J.~P., {Brown},
  M.~A., \& {Jao}, W.-C. 2005{\natexlab{b}}, \aj, 129, 409

\bibitem[{{Deacon} {et~al.}(2009{\natexlab{a}}){Deacon}, {Hambly}, {King}, \&
  {McCaughrean}}]{Deacon(2009a)}
{Deacon}, N.~R., {Hambly}, N.~C., {King}, R.~R., \& {McCaughrean}, M.~J.
  2009{\natexlab{a}}, \mnras, 394, 857

\bibitem[{{Deacon} {et~al.}(2009{\natexlab{b}}){Deacon}, {Groot}, {Drew},
  {Greimel}, {Hambly}, {Irwin}, {Aungwerojwit}, {Drake}, \&
  {Steeghs}}]{Deacon(2009b)}
{Deacon}, N.~R., {et~al.} 2009{\natexlab{b}}, \mnras, 397, 1685

\bibitem[{{Dieterich} {et~al.}(2012){Dieterich}, {Henry}, {Golimowski},
  {Krist}, \& {Tanner}}]{Dieterich(2012)}
{Dieterich}, S.~B., {Henry}, T.~J., {Golimowski}, D.~A., {Krist}, J.~E., \&
  {Tanner}, A.~M. 2012, \aj, 144, 64

\bibitem[{{Dieterich} {et~al.}(2014){Dieterich}, {Henry}, {Jao}, {Winters},
  {Hosey}, {Riedel}, \& {Subasavage}}]{Dieterich(2014)}
{Dieterich}, S.~B., {Henry}, T.~J., {Jao}, W.-C., {Winters}, J.~G., {Hosey},
  A.~D., {Riedel}, A.~R., \& {Subasavage}, J.~P. 2014, \aj, 147, 94

\bibitem[{{Dupuy} \& {Liu}(2012)}]{Dupuy(2012)}
{Dupuy}, T.~J., \& {Liu}, M.~C. 2012, \apjs, 201, 19

\bibitem[{{Fabricius} \& {Makarov}(2000)}]{Fabricius(2000)}
{Fabricius}, C., \& {Makarov}, V.~V. 2000, \aaps, 144, 45

\bibitem[{{Faherty} {et~al.}(2012){Faherty}, {Burgasser}, {Walter}, {Van der
  Bliek}, {Shara}, {Cruz}, {West}, {Vrba}, \&
  {Anglada-Escud{\'e}}}]{Faherty(2012)}
{Faherty}, J.~K., {et~al.} 2012, \apj, 752, 56

\bibitem[{{Finch} {et~al.}(2007){Finch}, {Henry}, {Subasavage}, {Jao}, \&
  {Hambly}}]{Finch(2007)}
{Finch}, C.~T., {Henry}, T.~J., {Subasavage}, J.~P., {Jao}, W.-C., \& {Hambly},
  N.~C. 2007, \aj, 133, 2898

\bibitem[{{Finch} {et~al.}(2012){Finch}, {Zacharias}, {Boyd}, {Henry}, \&
  {Hambly}}]{Finch(2012)}
{Finch}, C.~T., {Zacharias}, N., {Boyd}, M.~R., {Henry}, T.~J., \& {Hambly},
  N.~C. 2012, \apj, 745, 118

\bibitem[{{Finch} {et~al.}(2010){Finch}, {Zacharias}, \& {Henry}}]{Finch(2010)}
{Finch}, C.~T., {Zacharias}, N., \& {Henry}, T.~J. 2010, \aj, 140, 844

\bibitem[{{Fischer} \& {Marcy}(1992)}]{Fischer(1992)}
{Fischer}, D.~A., \& {Marcy}, G.~W. 1992, \apj, 396, 178

\bibitem[{{Gatewood} {et~al.}(2003){Gatewood}, {Coban}, \&
  {Han}}]{Gatewood(2003)}
{Gatewood}, G., {Coban}, L., \& {Han}, I. 2003, \aj, 125, 1530

\bibitem[{{Giclas} {et~al.}(1971){Giclas}, {Burnham}, \&
  {Thomas}}]{Giclas(1971)}
{Giclas}, H.~L., {Burnham}, R., \& {Thomas}, N.~G. 1971, {Lowell proper motion
  survey Northern Hemisphere. The G numbered stars. 8991 stars fainter than
  magnitude 8 with motions {gt} 0''.26/year} (Lowell Observatory, Flagstaff,
  Arizona)

\bibitem[{{Giclas} {et~al.}(1978){Giclas}, {Burnham}, \&
  {Thomas}}]{Giclas(1978)}
{Giclas}, H.~L., {Burnham}, Jr., R., \& {Thomas}, N.~G. 1978, Lowell
  Observatory Bulletin, 8, 89

\bibitem[{{Gizis}(1997)}]{Gizis(1997)}
{Gizis}, J.~E. 1997, \aj, 113, 806

\bibitem[{{Gizis} {et~al.}(2011){Gizis}, {Troup}, \& {Burgasser}}]{Gizis(2011)}
{Gizis}, J.~E., {Troup}, N.~W., \& {Burgasser}, A.~J. 2011, \apjl, 736, L34

\bibitem[{{Graham}(1982)}]{Graham(1982)}
{Graham}, J.~A. 1982, \pasp, 94, 244

\bibitem[{{Hambly} {et~al.}(2004){Hambly}, {Henry}, {Subasavage}, {Brown}, \&
  {Jao}}]{Hambly(2004)}
{Hambly}, N.~C., {Henry}, T.~J., {Subasavage}, J.~P., {Brown}, M.~A., \& {Jao},
  W.-C. 2004, \aj, 128, 437

\bibitem[{{Hambly} {et~al.}(2001){Hambly}, {Irwin}, \&
  {MacGillivray}}]{Hambly(2001)}
{Hambly}, N.~C., {Irwin}, M.~J., \& {MacGillivray}, H.~T. 2001, \mnras, 326,
  1295

\bibitem[{{Hawley} {et~al.}(1996){Hawley}, {Gizis}, \& {Reid}}]{Hawley(1996)}
{Hawley}, S.~L., {Gizis}, J.~E., \& {Reid}, I.~N. 1996, \aj, 112, 2799

\bibitem[{{Henry}(1991)}]{Henry(1991)}
{Henry}, T.~J. 1991, PhD thesis, Arizona Univ., Tucson.

\bibitem[{Henry(2014)}]{Henry(2014)}
Henry, T.~J. 2014, in The Observer's Handbook, ed. D.~M.~F. Chapman (Toronto,
  Ontario: The Royal Astronomical Society of Canada), 286--290

\bibitem[{{Henry} {et~al.}(1997){Henry}, {Ianna}, {Kirkpatrick}, \&
  {Jahreiss}}]{Henry(1997)}
{Henry}, T.~J., {Ianna}, P.~A., {Kirkpatrick}, J.~D., \& {Jahreiss}, H. 1997,
  \aj, 114, 388

\bibitem[{{Henry} {et~al.}(2006){Henry}, {Jao}, {Subasavage}, {Beaulieu},
  {Ianna}, {Costa}, \& {M{\'e}ndez}}]{Henry(2006)}
{Henry}, T.~J., {Jao}, W.-C., {Subasavage}, J.~P., {Beaulieu}, T.~D., {Ianna},
  P.~A., {Costa}, E., \& {M{\'e}ndez}, R.~A. 2006, \aj, 132, 2360

\bibitem[{{Henry} \& {McCarthy}(1990)}]{Henry(1990)}
{Henry}, T.~J., \& {McCarthy}, Jr., D.~W. 1990, \apj, 350, 334

\bibitem[{{Henry} \& {McCarthy}(1993)}]{Henry(1993)}
---. 1993, \aj, 106, 773

\bibitem[{{Henry} {et~al.}(2004){Henry}, {Subasavage}, {Brown}, {Beaulieu},
  {Jao}, \& {Hambly}}]{Henry(2004)}
{Henry}, T.~J., {Subasavage}, J.~P., {Brown}, M.~A., {Beaulieu}, T.~D., {Jao},
  W.-C., \& {Hambly}, N.~C. 2004, \aj, 128, 2460

\bibitem[{{Hershey} \& {Taff}(1998)}]{Hershey(1998)}
{Hershey}, J.~L., \& {Taff}, L.~G. 1998, \aj, 116, 1440

\bibitem[{{H{\o}g} {et~al.}(2000){H{\o}g}, {Fabricius}, {Makarov}, {Urban},
  {Corbin}, {Wycoff}, {Bastian}, {Schwekendiek}, \& {Wicenec}}]{Hog(2000)}
{H{\o}g}, E., {et~al.} 2000, \aap, 355, L27

\bibitem[{{Janson} {et~al.}(2014){Janson}, {Bergfors}, {Brandner},
  {Kudryavtseva}, {Hormuth}, {Hippler}, \& {Henning}}]{Janson(2014)}
{Janson}, M., {Bergfors}, C., {Brandner}, W., {Kudryavtseva}, N., {Hormuth},
  F., {Hippler}, S., \& {Henning}, T. 2014, ArXiv e-prints

\bibitem[{{Janson} {et~al.}(2012){Janson}, {Hormuth}, {Bergfors}, {Brandner},
  {Hippler}, {Daemgen}, {Kudryavtseva}, {Schmalzl}, {Schnupp}, \&
  {Henning}}]{Janson(2012)}
{Janson}, M., {et~al.} 2012, \apj, 754, 44

\bibitem[{{Jao} {et~al.}(2008){Jao}, {Henry}, {Beaulieu}, \&
  {Subasavage}}]{Jao(2008)}
{Jao}, W.-C., {Henry}, T.~J., {Beaulieu}, T.~D., \& {Subasavage}, J.~P. 2008,
  \aj, 136, 840

\bibitem[{{Jao} {et~al.}(2005){Jao}, {Henry}, {Subasavage}, {Brown}, {Ianna},
  {Bartlett}, {Costa}, \& {M{\'e}ndez}}]{Jao(2005)}
{Jao}, W.-C., {Henry}, T.~J., {Subasavage}, J.~P., {Brown}, M.~A., {Ianna},
  P.~A., {Bartlett}, J.~L., {Costa}, E., \& {M{\'e}ndez}, R.~A. 2005, \aj, 129,
  1954

\bibitem[{{Jao} {et~al.}(2014){Jao}, {Henry}, {Subasavage}, {Winters}, {Gies},
  {Riedel}, \& {Ianna}}]{Jao(2014)}
{Jao}, W.-C., {Henry}, T.~J., {Subasavage}, J.~P., {Winters}, J.~G., {Gies},
  D.~R., {Riedel}, A.~R., \& {Ianna}, P.~A. 2014, \aj, 147, 21

\bibitem[{{Jao} {et~al.}(2011){Jao}, {Henry}, {Subasavage}, {Winters},
  {Riedel}, \& {Ianna}}]{Jao(2011)}
{Jao}, W.-C., {Henry}, T.~J., {Subasavage}, J.~P., {Winters}, J.~G., {Riedel},
  A.~R., \& {Ianna}, P.~A. 2011, \aj, 141, 117

\bibitem[{{Kilkenny} \& {Cousins}(1995)}]{Kilkenny(1995)}
{Kilkenny}, D., \& {Cousins}, A.~W.~J. 1995, \apss, 230, 155

\bibitem[{{Kilkenny} {et~al.}(2007){Kilkenny}, {Koen}, {van Wyk}, {Marang}, \&
  {Cooper}}]{Kilkenny(2007)}
{Kilkenny}, D., {Koen}, C., {van Wyk}, F., {Marang}, F., \& {Cooper}, D. 2007,
  \mnras, 380, 1261

\bibitem[{{Kilkenny} {et~al.}(1998){Kilkenny}, {van Wyk}, {Roberts}, {Marang},
  \& {Cooper}}]{Kilkenny(1998)}
{Kilkenny}, D., {van Wyk}, F., {Roberts}, G., {Marang}, F., \& {Cooper}, D.
  1998, \mnras, 294, 93

\bibitem[{{Koen} {et~al.}(2002){Koen}, {Kilkenny}, {van Wyk}, {Cooper}, \&
  {Marang}}]{Koen(2002)}
{Koen}, C., {Kilkenny}, D., {van Wyk}, F., {Cooper}, D., \& {Marang}, F. 2002,
  \mnras, 334, 20

\bibitem[{{Koen} {et~al.}(2010){Koen}, {Kilkenny}, {van Wyk}, \&
  {Marang}}]{Koen(2010)}
{Koen}, C., {Kilkenny}, D., {van Wyk}, F., \& {Marang}, F. 2010, \mnras, 403,
  1949

\bibitem[{{Landolt}(1992)}]{Landolt(1992)}
{Landolt}, A.~U. 1992, \aj, 104, 372

\bibitem[{{Landolt}(2007)}]{Landolt(2007)}
---. 2007, \aj, 133, 2502

\bibitem[{{Landolt}(2013)}]{Landolt(2013)}
---. 2013, \aj, 146, 131

\bibitem[{{L{\'e}pine}(2005{\natexlab{a}})}]{Lepine(2005b)}
{L{\'e}pine}, S. 2005{\natexlab{a}}, \aj, 130, 1680

\bibitem[{{L{\'e}pine}(2005{\natexlab{b}})}]{Lepine(2005a)}
---. 2005{\natexlab{b}}, \aj, 130, 1247

\bibitem[{{L{\'e}pine}(2008)}]{Lepine(2008)}
---. 2008, \aj, 135, 2177

\bibitem[{{L{\'e}pine} \& {Gaidos}(2011)}]{Lepine(2011)}
{L{\'e}pine}, S., \& {Gaidos}, E. 2011, \aj, 142, 138

\bibitem[{{L{\'e}pine} {et~al.}(2002){L{\'e}pine}, {Shara}, \&
  {Rich}}]{Lepine(2002)}
{L{\'e}pine}, S., {Shara}, M.~M., \& {Rich}, R.~M. 2002, \aj, 124, 1190

\bibitem[{{L{\'e}pine} {et~al.}(2003){L{\'e}pine}, {Shara}, \&
  {Rich}}]{Lepine(2003)}
---. 2003, \aj, 126, 921

\bibitem[{{Luyten}(1979{\natexlab{a}})}]{Luyten(1979a)}
{Luyten}, W.~J. 1979{\natexlab{a}}, {LHS Catalogue. Second edition.}
  (University of Minnesota)

\bibitem[{{Luyten}(1979{\natexlab{b}})}]{Luyten(1979b)}
---. 1979{\natexlab{b}}, {NLTT catalogue. Volume\_I. +90\_\_to\_+30\_.
  Volume.\_II. +30\_\_to\_0\_.} (University of Minnesota)

\bibitem[{{Luyten}(1980{\natexlab{a}})}]{Luyten(1980a)}
---. 1980{\natexlab{a}}, {NLTT Catalogue. Volume\_III. 0\_\_to -30\_.}
  (University of Minnesota)

\bibitem[{{Luyten}(1980{\natexlab{b}})}]{Luyten(1980b)}
---. 1980{\natexlab{b}}, {NLTT Catalogue. Volume\_IV. -30\_\_to\_-90\_.}
  (University of Minnesota)

\bibitem[{{Mamajek} {et~al.}(2013){Mamajek}, {Bartlett}, {Seifahrt}, {Henry},
  {Dieterich}, {Lurie}, {Kenworthy}, {Jao}, {Riedel}, {Subasavage}, {Winters},
  {Finch}, {Ianna}, \& {Bean}}]{Mamajek(2013)}
{Mamajek}, E.~E., {et~al.} 2013, \aj, 146, 154

\bibitem[{{Monet} {et~al.}(2003){Monet}, {Levine}, {Canzian}, {Ables}, {Bird},
  {Dahn}, {Guetter}, {Harris}, {Henden}, {Leggett}, {Levison}, {Luginbuhl},
  {Martini}, {Monet}, {Munn}, {Pier}, {Rhodes}, {Riepe}, {Sell}, {Stone},
  {Vrba}, {Walker}, {Westerhout}, {Brucato}, {Reid}, {Schoening}, {Hartley},
  {Read}, \& {Tritton}}]{Monet(2003)}
{Monet}, D.~G., {et~al.} 2003, \aj, 125, 984

\bibitem[{{Morgan}(1995)}]{Morgan(1995)}
{Morgan}, D.~H. 1995, in Astronomical Society of the Pacific Conference Series,
  Vol.~84, IAU Colloq. 148: The Future Utilisation of Schmidt Telescopes, ed.
  J.~{Chapman}, R.~{Cannon}, S.~{Harrison}, \& B.~{Hidayat}, 137

\bibitem[{{Patterson} {et~al.}(1998){Patterson}, {Ianna}, \&
  {Begam}}]{Patterson(1998)}
{Patterson}, R.~J., {Ianna}, P.~A., \& {Begam}, M.~C. 1998, \aj, 115, 1648

\bibitem[{{Perryman} {et~al.}(1997){Perryman}, {Lindegren}, {Kovalevsky},
  {Hoeg}, {Bastian}, {Bernacca}, {Cr{\'e}z{\'e}}, {Donati}, {Grenon},
  {Grewing}, {van Leeuwen}, {van der Marel}, {Mignard}, {Murray}, {Le Poole},
  {Schrijver}, {Turon}, {Arenou}, {Froeschl{\'e}}, \&
  {Petersen}}]{Perryman(1997)}
{Perryman}, M.~A.~C., {et~al.} 1997, \aap, 323, L49

\bibitem[{{Phan-Bao}(2011)}]{Phan-Bao(2011)}
{Phan-Bao}, N. 2011, Astronomische Nachrichten, 332, 668

\bibitem[{{Pokorny} {et~al.}(2003){Pokorny}, {Jones}, \&
  {Hambly}}]{Pokorny(2003)}
{Pokorny}, R.~S., {Jones}, H.~R.~A., \& {Hambly}, N.~C. 2003, \aap, 397, 575

\bibitem[{{Pokorny} {et~al.}(2004){Pokorny}, {Jones}, {Hambly}, \&
  {Pinfield}}]{Pokorny(2004)}
{Pokorny}, R.~S., {Jones}, H.~R.~A., {Hambly}, N.~C., \& {Pinfield}, D.~J.
  2004, \aap, 421, 763

\bibitem[{{Reid} \& {Gizis}(2005)}]{Reid(2005)}
{Reid}, I.~N., \& {Gizis}, J.~E. 2005, \pasp, 117, 676

\bibitem[{{Reid} {et~al.}(2002){Reid}, {Kilkenny}, \& {Cruz}}]{Reid(2002)}
{Reid}, I.~N., {Kilkenny}, D., \& {Cruz}, K.~L. 2002, \aj, 123, 2822

\bibitem[{{Reid} {et~al.}(2001){Reid}, {van Wyk}, {Marang}, {Roberts},
  {Kilkenny}, \& {Mahoney}}]{Reid(2001)}
{Reid}, I.~N., {van Wyk}, F., {Marang}, F., {Roberts}, G., {Kilkenny}, D., \&
  {Mahoney}, S. 2001, \mnras, 325, 931

\bibitem[{{Reid} {et~al.}(2003){Reid}, {Cruz}, {Allen}, {Mungall}, {Kilkenny},
  {Liebert}, {Hawley}, {Fraser}, {Covey}, \& {Lowrance}}]{Reid(2003)}
{Reid}, I.~N., {et~al.} 2003, \aj, 126, 3007

\bibitem[{{Reid} {et~al.}(2004){Reid}, {Cruz}, {Allen}, {Mungall}, {Kilkenny},
  {Liebert}, {Hawley}, {Fraser}, {Covey}, {Lowrance}, {Kirkpatrick}, \&
  {Burgasser}}]{Reid(2004)}
---. 2004, \aj, 128, 463

\bibitem[{{Riedel}(2012)}]{Riedel(2012)}
{Riedel}, A.~R. 2012, PhD thesis, Georgia State University

\bibitem[{{Riedel} {et~al.}(2011){Riedel}, {Murphy}, {Henry}, {Melis}, {Jao},
  \& {Subasavage}}]{Riedel(2011)}
{Riedel}, A.~R., {Murphy}, S.~J., {Henry}, T.~J., {Melis}, C., {Jao}, W.-C., \&
  {Subasavage}, J.~P. 2011, \aj, 142, 104

\bibitem[{{Riedel} {et~al.}(2010){Riedel}, {Subasavage}, {Finch}, {Jao},
  {Henry}, {Winters}, {Brown}, {Ianna}, {Costa}, \& {Mendez}}]{Riedel(2010)}
{Riedel}, A.~R., {et~al.} 2010, \aj, 140, 897

\bibitem[{{Riedel} {et~al.}(2014){Riedel}, {Finch}, {Henry}, {Subasavage},
  {Jao}, {Malo}, {Rodriguez}, {White}, {Gies}, {Dieterich}, {Winters},
  {Davison}, {Nelan}, {Blunt}, {Cruz}, {Rice}, \& {Ianna}}]{Riedel(2014)}
---. 2014, \aj, 147, 85

\bibitem[{{Ruiz} {et~al.}(1993){Ruiz}, {Takamiya}, {Mendez}, {Maza}, \&
  {Wishnjewsky}}]{Ruiz(1993)}
{Ruiz}, M.~T., {Takamiya}, M.~Y., {Mendez}, R., {Maza}, J., \& {Wishnjewsky},
  M. 1993, \aj, 106, 2575

\bibitem[{{Ruiz} {et~al.}(2001){Ruiz}, {Wischnjewsky}, {Rojo}, \&
  {Gonzalez}}]{Ruiz(2001)}
{Ruiz}, M.~T., {Wischnjewsky}, M., {Rojo}, P.~M., \& {Gonzalez}, L.~E. 2001,
  \apjs, 133, 119

\bibitem[{{Schilbach} {et~al.}(2009){Schilbach}, {R{\"o}ser}, \&
  {Scholz}}]{Schilbach(2009)}
{Schilbach}, E., {R{\"o}ser}, S., \& {Scholz}, R.-D. 2009, \aap, 493, L27

\bibitem[{{Schmidt} {et~al.}(2007){Schmidt}, {Cruz}, {Bongiorno}, {Liebert}, \&
  {Reid}}]{Schmidt(2007)}
{Schmidt}, S.~J., {Cruz}, K.~L., {Bongiorno}, B.~J., {Liebert}, J., \& {Reid},
  I.~N. 2007, \aj, 133, 2258

\bibitem[{{Scholz} {et~al.}(2004){Scholz}, {Lehmann}, {Matute}, \&
  {Zinnecker}}]{Scholz(2004)}
{Scholz}, R.-D., {Lehmann}, I., {Matute}, I., \& {Zinnecker}, H. 2004, \aap,
  425, 519

\bibitem[{{Shkolnik} {et~al.}(2012){Shkolnik}, {Anglada-Escud{\'e}}, {Liu},
  {Bowler}, {Weinberger}, {Boss}, {Reid}, \& {Tamura}}]{Shkolnik(2012)}
{Shkolnik}, E.~L., {Anglada-Escud{\'e}}, G., {Liu}, M.~C., {Bowler}, B.~P.,
  {Weinberger}, A.~J., {Boss}, A.~P., {Reid}, I.~N., \& {Tamura}, M. 2012,
  \apj, 758, 56

\bibitem[{{Skrutskie} {et~al.}(2006){Skrutskie}, {Cutri}, {Stiening},
  {Weinberg}, {Schneider}, {Carpenter}, {Beichman}, {Capps}, {Chester},
  {Elias}, {Huchra}, {Liebert}, {Lonsdale}, {Monet}, {Price}, {Seitzer},
  {Jarrett}, {Kirkpatrick}, {Gizis}, {Howard}, {Evans}, {Fowler}, {Fullmer},
  {Hurt}, {Light}, {Kopan}, {Marsh}, {McCallon}, {Tam}, {Van Dyk}, \&
  {Wheelock}}]{Skrutskie(2006)}
{Skrutskie}, M.~F., {et~al.} 2006, \aj, 131, 1163

\bibitem[{{Smart} {et~al.}(2010){Smart}, {Ioannidis}, {Jones}, {Bucciarelli},
  \& {Lattanzi}}]{Smart(2010)}
{Smart}, R.~L., {Ioannidis}, G., {Jones}, H.~R.~A., {Bucciarelli}, B., \&
  {Lattanzi}, M.~G. 2010, \aap, 514, A84

\bibitem[{{Smart} {et~al.}(2007){Smart}, {Lattanzi}, {Jahrei{\ss}},
  {Bucciarelli}, \& {Massone}}]{Smart(2007)}
{Smart}, R.~L., {Lattanzi}, M.~G., {Jahrei{\ss}}, H., {Bucciarelli}, B., \&
  {Massone}, G. 2007, \aap, 464, 787

\bibitem[{{S{\"o}derhjelm}(1999)}]{Soderhjelm(1999)}
{S{\"o}derhjelm}, S. 1999, \aap, 341, 121

\bibitem[{{Sozzetti} {et~al.}(2014){Sozzetti}, {Giacobbe}, {Lattanzi},
  {Micela}, {Morbidelli}, \& {Tinetti}}]{Sozzetti(2014)}
{Sozzetti}, A., {Giacobbe}, P., {Lattanzi}, M.~G., {Micela}, G., {Morbidelli},
  R., \& {Tinetti}, G. 2014, \mnras, 437, 497

\bibitem[{{Subasavage} {et~al.}(2005{\natexlab{a}}){Subasavage}, {Henry},
  {Hambly}, {Brown}, \& {Jao}}]{Subasavage(2005a)}
{Subasavage}, J.~P., {Henry}, T.~J., {Hambly}, N.~C., {Brown}, M.~A., \& {Jao},
  W.-C. 2005{\natexlab{a}}, \aj, 129, 413

\bibitem[{{Subasavage} {et~al.}(2005{\natexlab{b}}){Subasavage}, {Henry},
  {Hambly}, {Brown}, {Jao}, \& {Finch}}]{Subasavage(2005b)}
{Subasavage}, J.~P., {Henry}, T.~J., {Hambly}, N.~C., {Brown}, M.~A., {Jao},
  W.-C., \& {Finch}, C.~T. 2005{\natexlab{b}}, \aj, 130, 1658

\bibitem[{{Subasavage} {et~al.}(2009){Subasavage}, {Jao}, {Henry}, {Bergeron},
  {Dufour}, {Ianna}, {Costa}, \& {M{\'e}ndez}}]{Subasavage(2009)}
{Subasavage}, J.~P., {Jao}, W.-C., {Henry}, T.~J., {Bergeron}, P., {Dufour},
  P., {Ianna}, P.~A., {Costa}, E., \& {M{\'e}ndez}, R.~A. 2009, \aj, 137, 4547

\bibitem[{{Tinney}(1996)}]{Tinney(1996)}
{Tinney}, C.~G. 1996, \mnras, 281, 644

\bibitem[{{Tinney} {et~al.}(1995){Tinney}, {Reid}, {Gizis}, \&
  {Mould}}]{Tinney(1995)}
{Tinney}, C.~G., {Reid}, I.~N., {Gizis}, J., \& {Mould}, J.~R. 1995, \aj, 110,
  3014

\bibitem[{{van Altena} {et~al.}(1995){van Altena}, {Lee}, \&
  {Hoffleit}}]{vanAltena(1995)}
{van Altena}, W.~F., {Lee}, J.~T., \& {Hoffleit}, D. 1995, VizieR Online Data
  Catalog, 1174, 0

\bibitem[{{van Leeuwen}(2007)}]{vanLeeuwen(2007)}
{van Leeuwen}, F., ed. 2007, Astrophysics and Space Science Library, Vol. 350,
  {Hipparcos, the New Reduction of the Raw Data}

\bibitem[{{von Braun} {et~al.}(2011){von Braun}, {Boyajian}, {Kane}, {van
  Belle}, {Ciardi}, {L{\'o}pez-Morales}, {McAlister}, {Henry}, {Jao}, {Riedel},
  {Subasavage}, {Schaefer}, {ten Brummelaar}, {Ridgway}, {Sturmann},
  {Sturmann}, {Mazingue}, {Turner}, {Farrington}, {Goldfinger}, \&
  {Boden}}]{vonBraun(2011)}
{von Braun}, K., {et~al.} 2011, \apjl, 729, L26

\bibitem[{{Weis}(1984)}]{Weis(1984)}
{Weis}, E.~W. 1984, \apjs, 55, 289

\bibitem[{{Weis}(1986)}]{Weis(1986)}
---. 1986, \aj, 91, 626

\bibitem[{{Weis}(1987)}]{Weis(1987)}
---. 1987, \aj, 93, 451

\bibitem[{{Weis}(1988)}]{Weis(1988)}
---. 1988, \apss, 142, 223

\bibitem[{{Weis}(1991{\natexlab{a}})}]{Weis(1991b)}
---. 1991{\natexlab{a}}, \aj, 102, 1795

\bibitem[{{Weis}(1991{\natexlab{b}})}]{Weis(1991a)}
---. 1991{\natexlab{b}}, \aj, 101, 1882

\bibitem[{{Weis}(1993)}]{Weis(1993)}
---. 1993, \aj, 105, 1962

\bibitem[{{Weis}(1994)}]{Weis(1994)}
---. 1994, \aj, 107, 1135

\bibitem[{{Weis}(1996)}]{Weis(1996)}
---. 1996, \aj, 112, 2300

\bibitem[{{Weis}(1999)}]{Weis(1999)}
---. 1999, \aj, 117, 3021

\bibitem[{{Winters} {et~al.}(2011){Winters}, {Henry}, {Jao}, {Subasavage},
  {Finch}, \& {Hambly}}]{Winters(2011)}
{Winters}, J.~G., {Henry}, T.~J., {Jao}, W.-C., {Subasavage}, J.~P., {Finch},
  C.~T., \& {Hambly}, N.~C. 2011, \aj, 141, 21

\bibitem[{{Wolf} \& {Reinmuth}(1925)}]{Wolf(1925)}
{Wolf}, M., \& {Reinmuth}, K. 1925, Astronomische Nachrichten, 223, 231

\bibitem[{{Wroblewski} \& {Costa}(1999)}]{Wroblewski(1999)}
{Wroblewski}, H., \& {Costa}, E. 1999, \aaps, 139, 25

\bibitem[{{Wroblewski} \& {Costa}(2000)}]{Wroblewski(2000)}
---. 2000, \aaps, 142, 369

\bibitem[{{Wroblewski} \& {Costa}(2001)}]{Wroblewski(2001)}
---. 2001, \aap, 367, 725

\bibitem[{{Wroblewski} \& {Torres}(1989)}]{Wroblewski(1989)}
{Wroblewski}, H., \& {Torres}, C. 1989, \aaps, 78, 231

\bibitem[{{Wroblewski} \& {Torres}(1990)}]{Wroblewski(1990)}
---. 1990, \aaps, 83, 317

\bibitem[{{Wroblewski} \& {Torres}(1991)}]{Wroblewski(1991)}
---. 1991, \aaps, 91, 129

\bibitem[{{Wroblewski} \& {Torres}(1992)}]{Wroblewski(1992)}
---. 1992, \aaps, 92, 449

\bibitem[{{Wroblewski} \& {Torres}(1994)}]{Wroblewski(1994)}
---. 1994, \aaps, 105, 179

\bibitem[{{Wroblewski} \& {Torres}(1995)}]{Wroblewski(1995)}
---. 1995, \aaps, 110, 27

\bibitem[{{Wroblewski} \& {Torres}(1996)}]{Wroblewski(1996)}
---. 1996, \aaps, 115, 481

\bibitem[{{Wroblewski} \& {Torres}(1997)}]{Wroblewski(1997)}
---. 1997, \aaps, 122, 447

\bibitem[{{Wroblewski} \& {Torres}(1998)}]{Wroblewski(1998)}
---. 1998, \aaps, 128, 457

\end{thebibliography}

\clearpage
\begin{figure}
 \includegraphics[scale=0.70,angle=90]{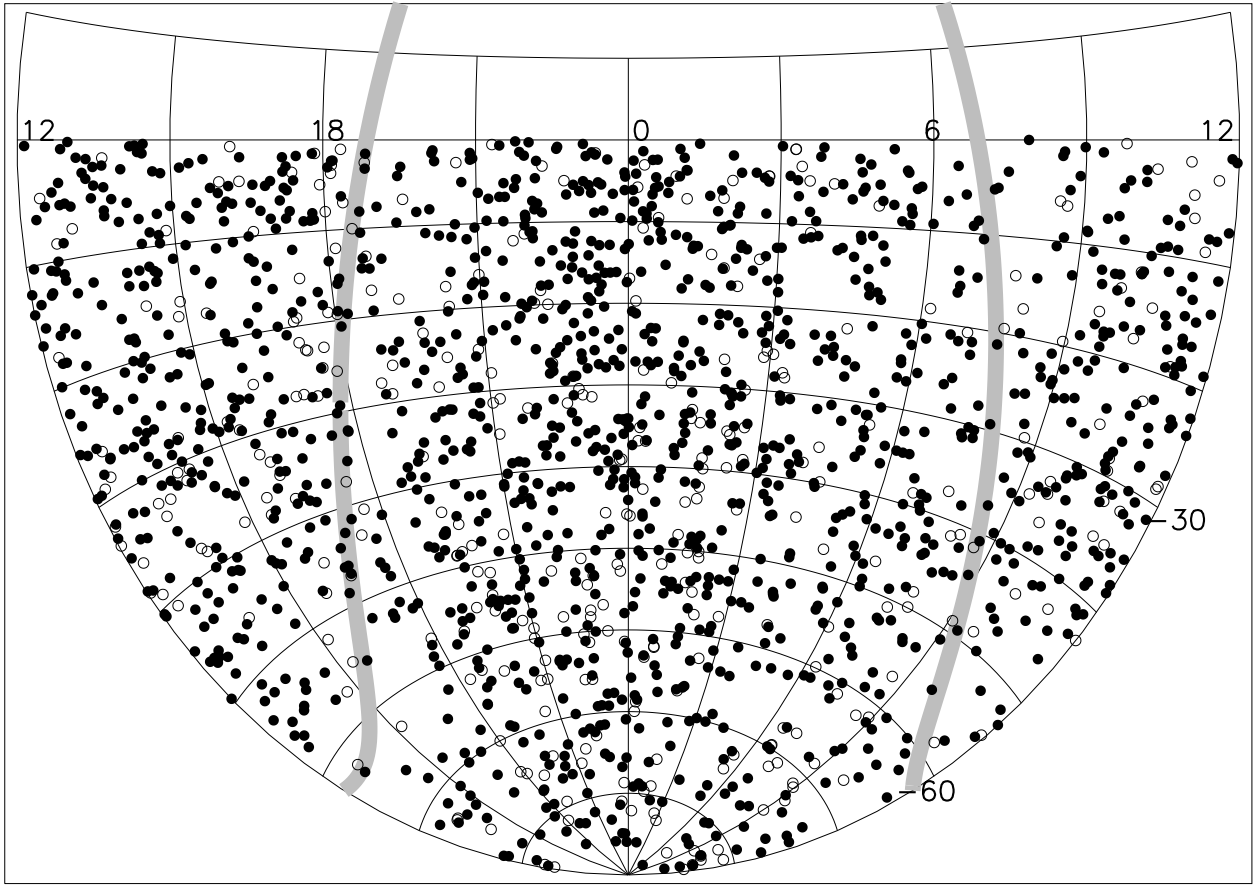}

\figcaption[fig1] {Distribution of the entire sample of 1748 M dwarf
  systems on the southern sky in RA and DEC.  Solid points indicate
  the 1404 systems within 25 pc, while open points indicate the 344
  systems beyond 25 pc.  The Galactic plane has been plotted in
  gray. \label{fig:winters1}}

\end{figure}

\begin{figure}
 \includegraphics[scale=0.70,angle=90]{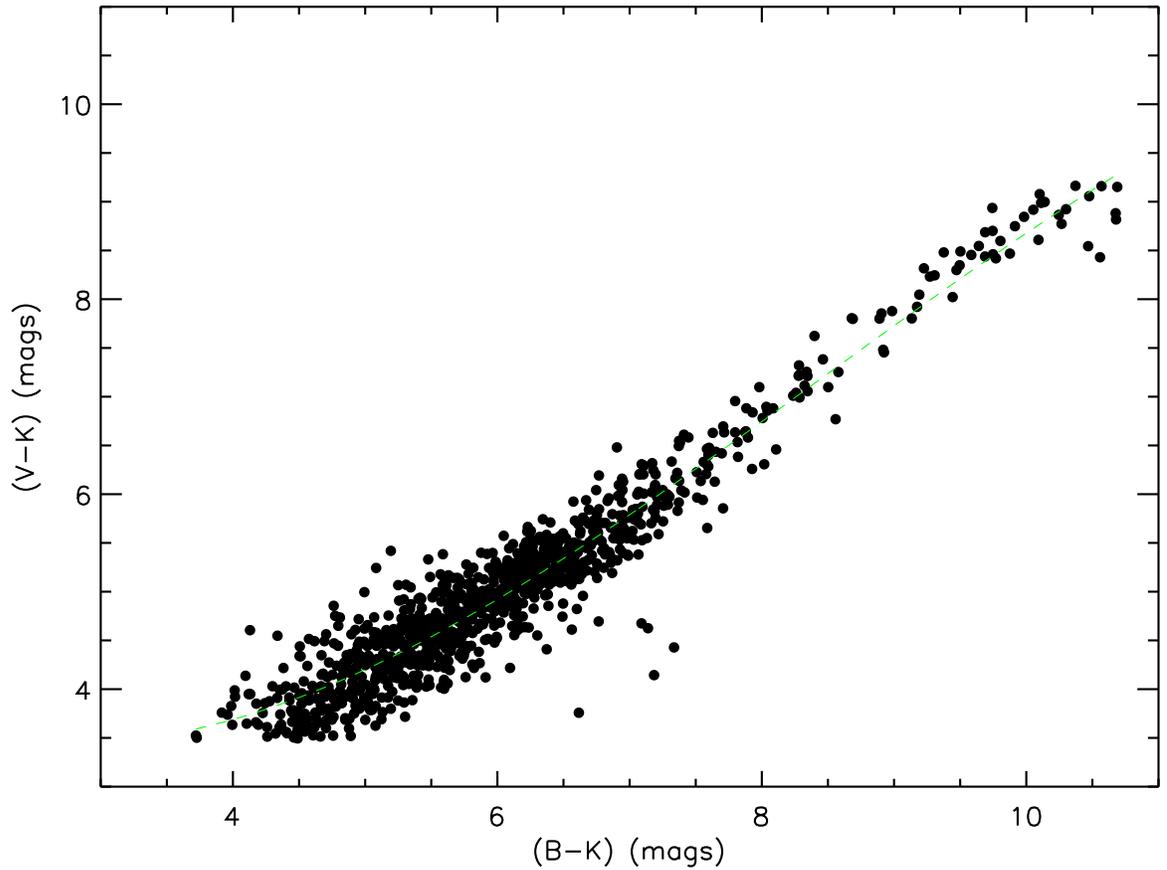}

\figcaption[fig2] {Comparison of the CCD ($V-K$) color to the plate
  ($B-K$) color for more than 1000 objects. Known multiples with
  blended photometry are not included. The dashed line indicates a
  third order polynomial fit (given in $\S3.2$) that provides an
  estimated CCD $V$ magnitude, given a measured plate $B$ magnitude
  for red objects and a known 2MASS $K$
  magnitude. \label{fig:winters2}}

\end{figure}

\begin{figure}
 \includegraphics[scale=0.70,angle=90]{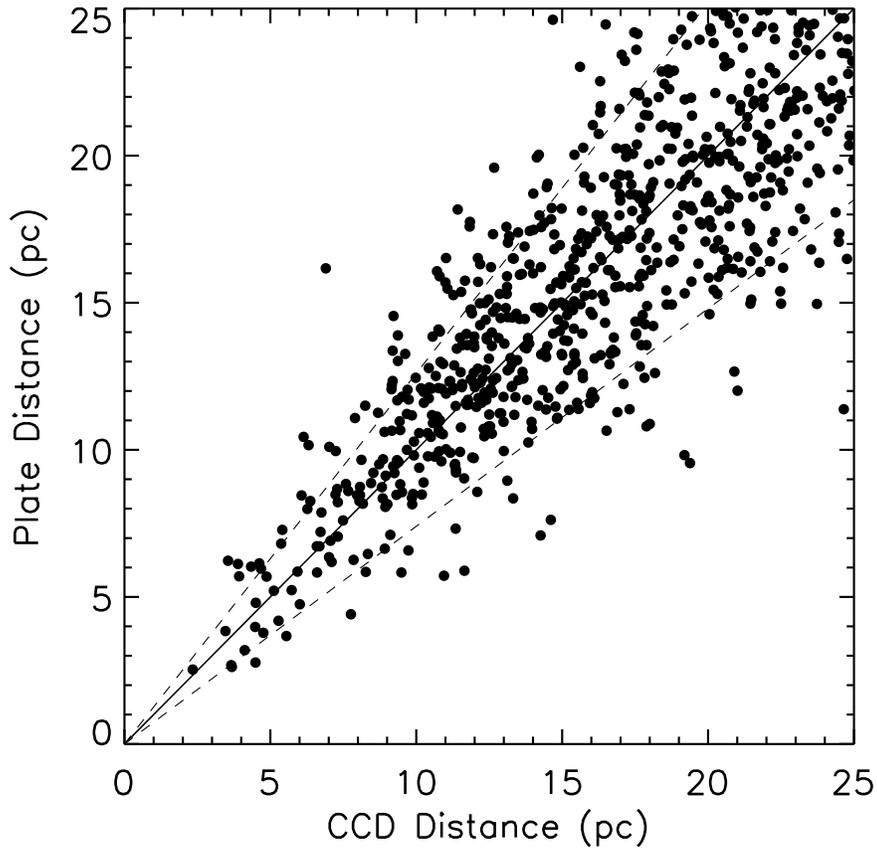}

\figcaption[fig3] {Distance comparisons of estimates using
  photographic plate photometry ({\it pltdist}) vs.~CCD photometry
  ({\it ccddist}) for the systems closer than 25 pc that have both
  $BRI$ and $VRI$ photometry.  Known unresolved multiples with blended
  photometry were not included.  The diagonal line represents 1:1
  correspondence in distances, while the dashed lines indicate the
  26\% errors associated with the plate distance
  estimates. \label{fig:winters3}}

\end{figure}
 
\begin{figure}
 \includegraphics[scale=0.70,angle=90]{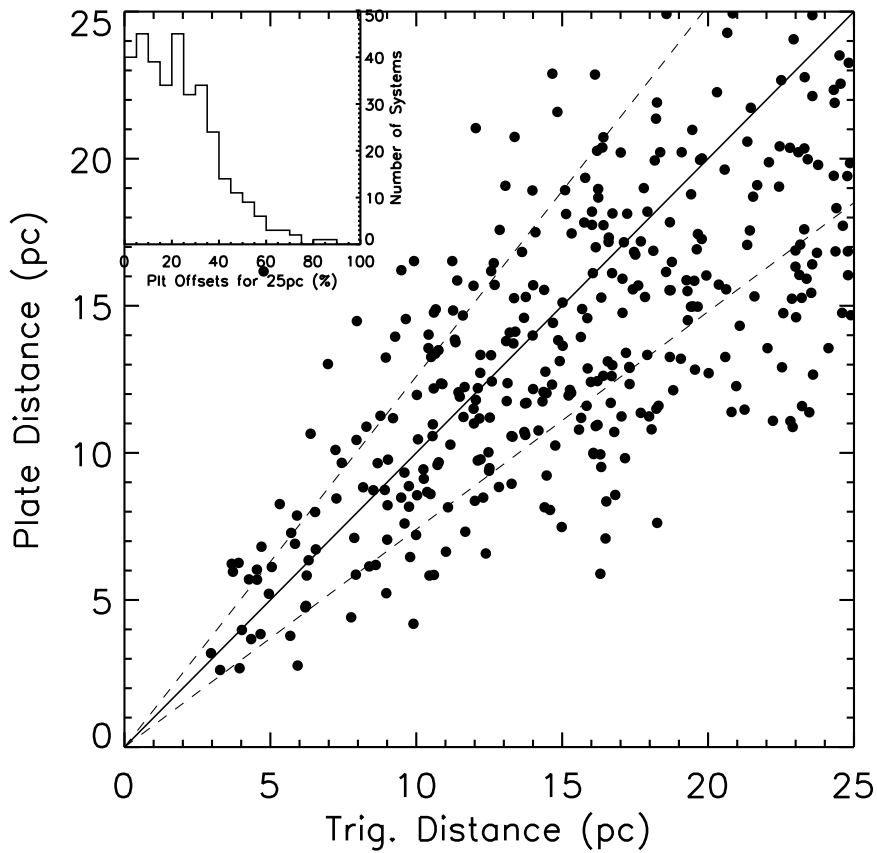}

\figcaption[fig4] {Distance comparisons of estimates from photographic
  plate photometry ({\it pltdist}) vs.~distances measured using
  trigonometric parallaxes ({\it trgdist}) for the systems closer than
  25 pc.  Known unresolved multiples with blended photometry were not
  included.  The diagonal line represents 1:1 correspondence in
  distances, while the dashed lines indicate the 26\% errors
  associated with the plate distance estimates. The inset histogram
  indicates the distribution of the distance offsets between the {\it
    pltdist} and {\it trgdist}. For this sample, the absolute mean
  offset is 24\%, consistent with the 26\% systematic error determined
  in \citet{Hambly(2004)}. \label{fig:winters4}}

\end{figure}
 
\begin{figure}
 \includegraphics[scale=0.70,angle=90]{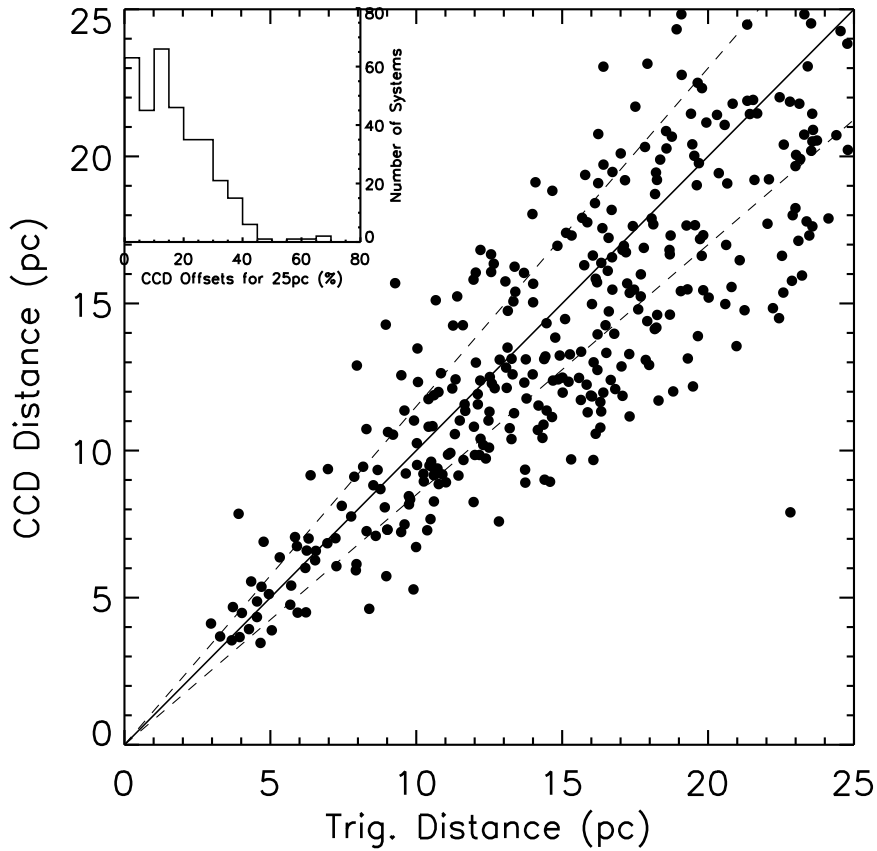}

\figcaption[fig5] {Distance comparisons of estimates from CCD
  photometry ({\it ccddist}) vs.~distances measured using
  trigonometric parallaxes ({\it trgdist}) for the systems closer than
  25 pc.  Known unresolved multiples with blended photometry were not
  included. The diagonal line represents 1:1 correspondence in
  distances, while the dashed lines indicate the 15\% errors
  associated with the CCD distance estimates.  Note the reduced
  scatter compared to the similar {\it pltdist} plot shown in Figure
  4, indicating the improvement in the photometry. The inset histogram
  indicates the distribution of the distance offsets between the {\it
    pltdist} and {\it trgdist}. For this sample, the absolute mean
  offset is 17\%, consistent with the 15\% systematic error determined
  by \citet{Henry(2004)}. \label{fig:winters5}}

\end{figure}
 
\begin{figure}
 \includegraphics[scale=0.70,angle=90]{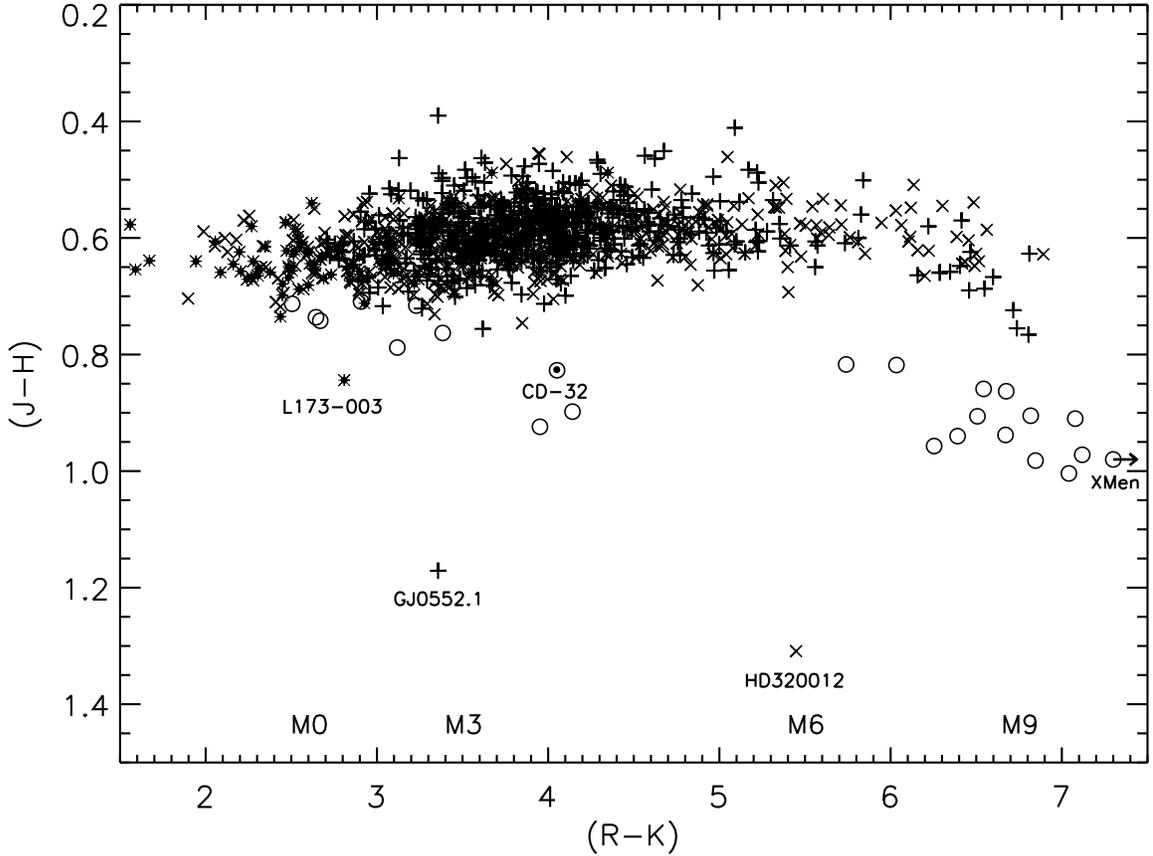}

\figcaption[fig6] {Color-color diagram showing ($J-H$) vs.~($R-K$)
  for all presumed single M dwarfs in the sample.  Known unresolved
  multiples with blended photometry were not included. Colors that use
  plate $R$ magnitudes are shown as x's, while those that use CCD $R$
  magnitudes are plotted as plusses.  Plate $R$ photometry is used to
  calculate colors for the objects for which a published parallax, but
  no CCD $R$ photometry exists and are noted as asterisks with a solid
  center. For comparison, a few known giants are denoted by open
  circles.  The object CD-32 16735 is plotted as a solid dot
  surrounded by an open circle, as it is surely a giant, given its
  preliminary parallax. The giant X Men is too red for the ($R-K$)
  color cut-off of this plot, but has been indicated with an
  arrow. Corresponding spectral type estimates are given along the
  bottom. \label{fig:winters6}}

\end{figure}

\begin{figure}
 \includegraphics[scale=0.70,angle=90]{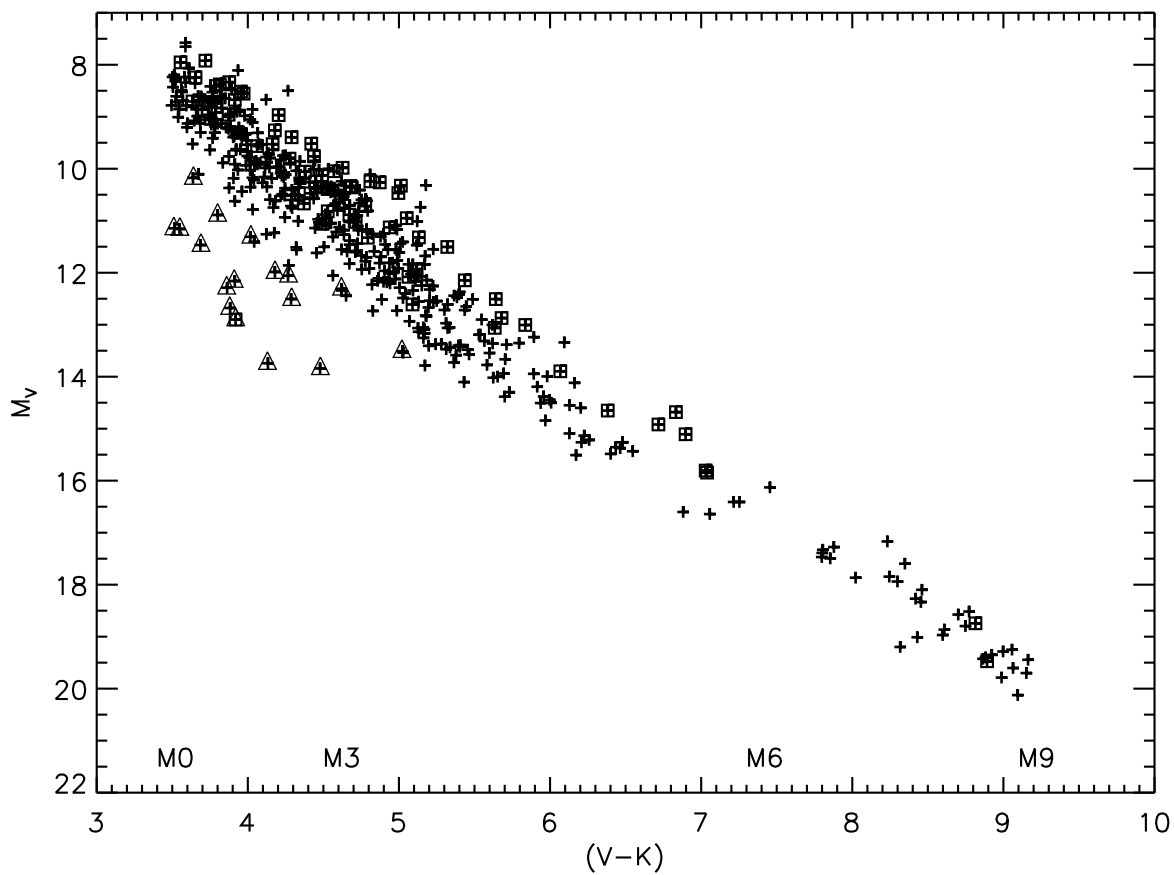}

\figcaption[fig7] {An observational HR diagram for all M dwarf systems
  in the sample that have trigonometric parallaxes and $V$ magnitudes,
  using $M_{V}$ and ($V-K$) as proxies for luminosity and temperature,
  respectively.  Known multiples with blended photometry are enclosed
  in open squares.  A set of a few dozen subdwarfs is evident below
  the main sequence. Spectroscopically confirmed subdwarfs have been
  enclosed in triangles, with references given in Table
  3. Corresponding spectral type estimates are given along the
  bottom. \label{fig:winters7}}

\end{figure}

\begin{figure}
 \includegraphics[scale=0.70,angle=90]{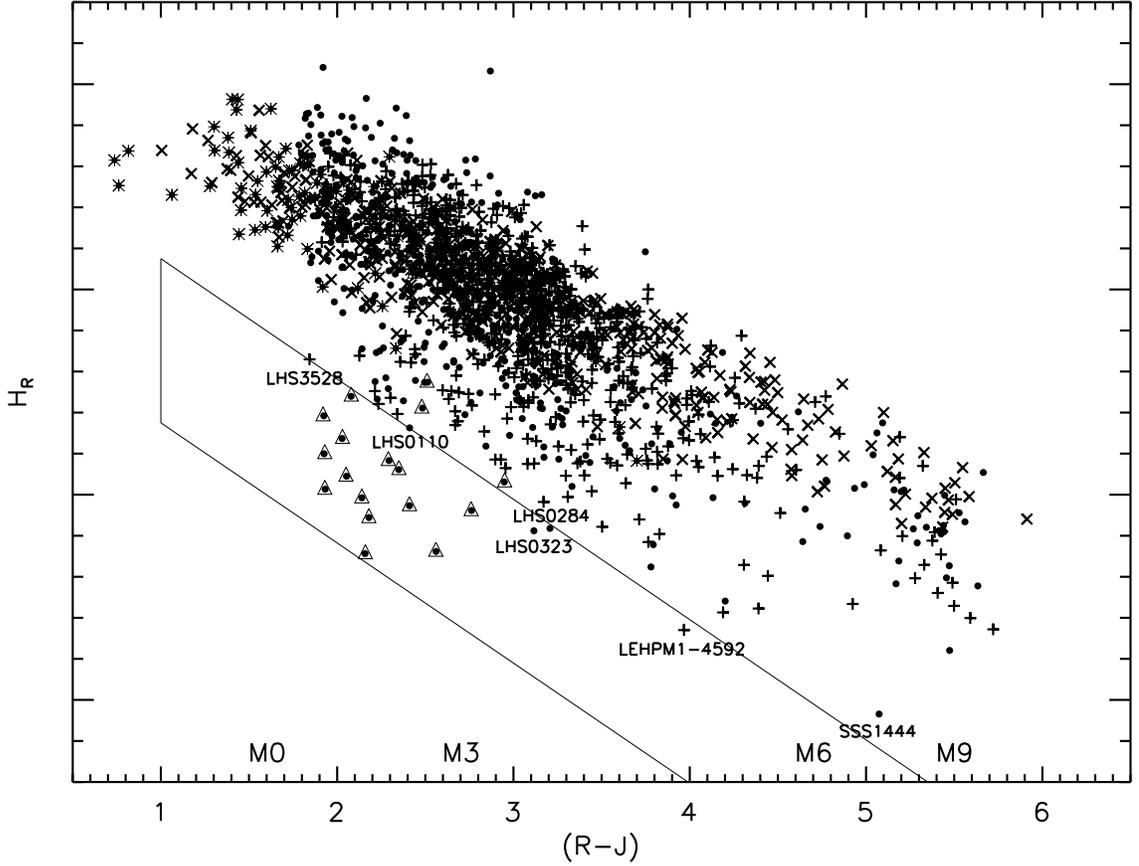}
 
\figcaption[fig8] {A reduced proper motion diagram is shown, which is
  used to separate the dwarfs from the subdwarfs in the entire sample,
  based on the proper motion of the objects. The region of interest is
  outlined, as in \citet{Boyd(2011a)}.  The y-axis is $H_R$ $=$ $R$ +
  5 + 5log~($\mu$) and the x-axis is the ($R-J$) color.  Primaries
  with {\it trgdists} are plotted as solid dots, those with {\it
    ccddists} are plusses, and those with {\it pltdists} are x's.
  Objects that have a published parallax, but no CCD photometry have
  been indicated as asterisks with a solid center. Three additional
  objects, among those identified as subdwarf candidates in Table 4,
  are shown inside the outlined subdwarf region: LHS0284, LHS0323, and
  LEHPM1-4592, although both LHS0284 and LHS0323 have been shown to be
  dwarfs (\citet{Jao(2011)}).  Spectroscopically confirmed subdwarfs
  have been enclosed in triangles, three of which fall outside the
  indicated subdwarf region and are labeled. Corresponding spectral
  type estimates are given along the bottom.
  \label{fig:winters8}}


\end{figure}
 
\begin{figure}
\includegraphics[scale=0.70,angle=90]{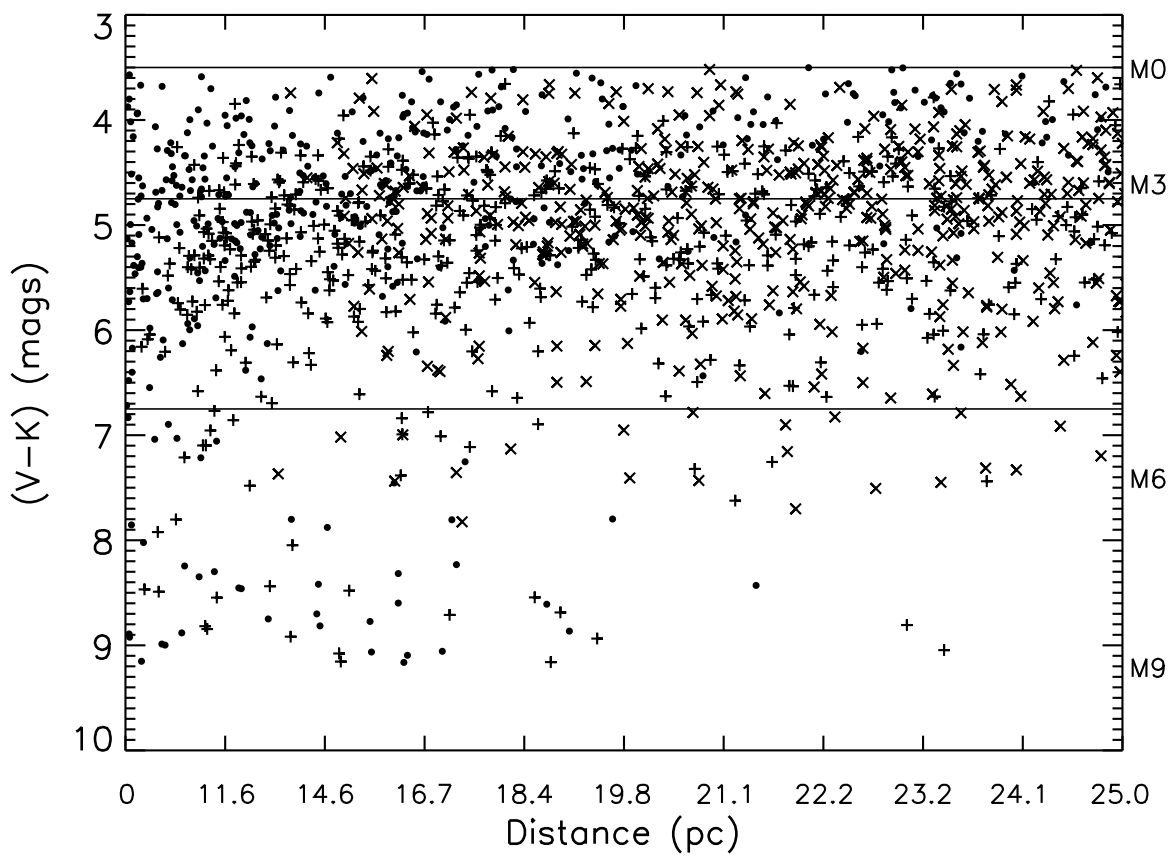}
 
\figcaption[fig9] {The density of M dwarf systems in the
  southern sky is plotted using ($V-K$) vs.~distance in 10
  equal-volume shells to 25 pc .  Solid points indicate systems with
  {\it trgdists}, plusses indicate those with {\it ccddists}, and x's
  indicate those with {\it pltdists}.  Corresponding spectral type
  estimates are given along the right.  For those 507 objects without
  a CCD $V$ magnitude, one has been estimated using the equation in
  $\S3.2$. The horizontal lines highlight masses of 0.60, 0.30, and
  0.10 M$_{\odot}$ from top to bottom. \label{fig:winters9}}

\end{figure}
 
\begin{figure}
\includegraphics[scale=0.70,angle=90]{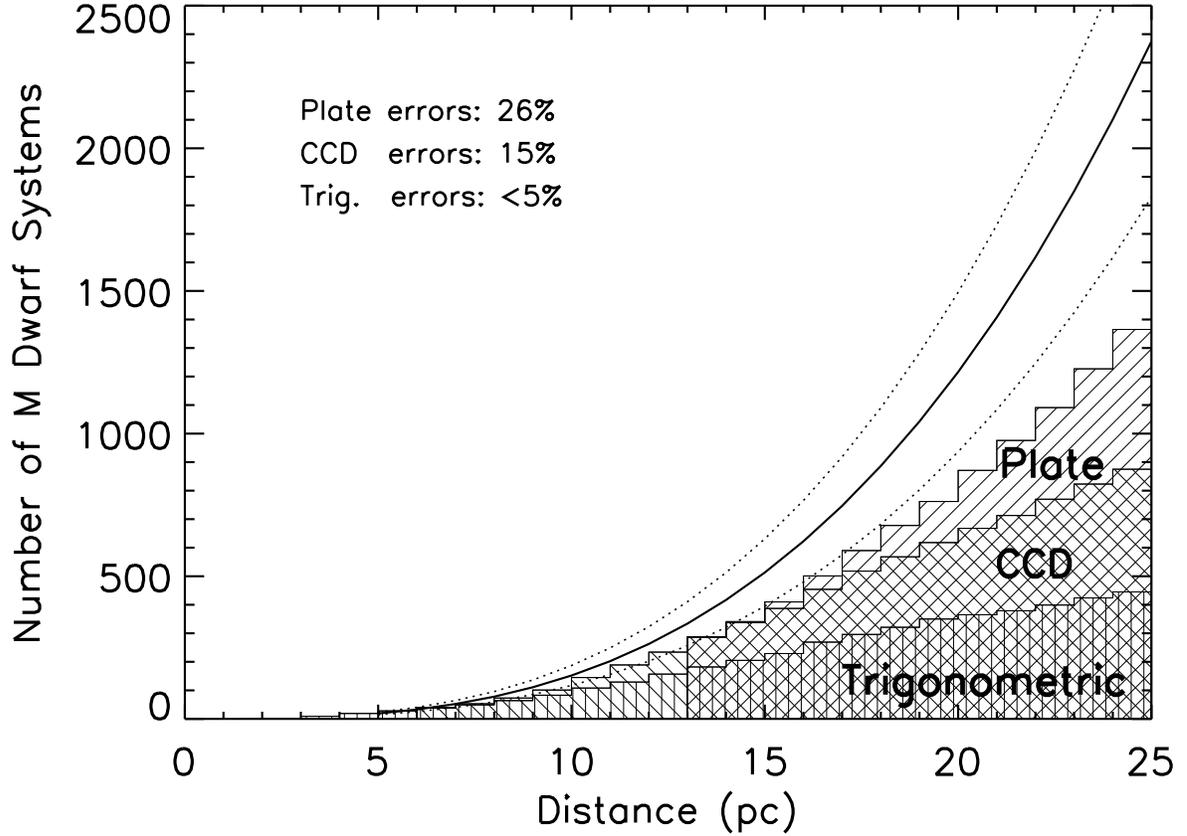}
 
\figcaption[Fig.10]{The cumulative numbers of southern
  single M dwarf systems known via the three distance techniques
  described in this paper are shown using histograms.  Those objects
  with {\it trgdists} are indicated with vertical hatching, those with
  {\it ccddists} have $-$45 degree hatching and those with {\it
    pltdists} have $+$45 degree hatching. The solid black curve
  indicates the expected number within that distance horizon based on
  the 19 southern M dwarf systems known to lie within 5 pc with
  accurate trigonometric parallaxes and assuming a uniform density to
  25 pc. The dotted lines indicate the extrapolated 23\% Poisson
  errors based on those 19 systems.\label{fig:winters10}}

\end{figure}
 
\begin{figure}
 \includegraphics[scale=0.70,angle=90]{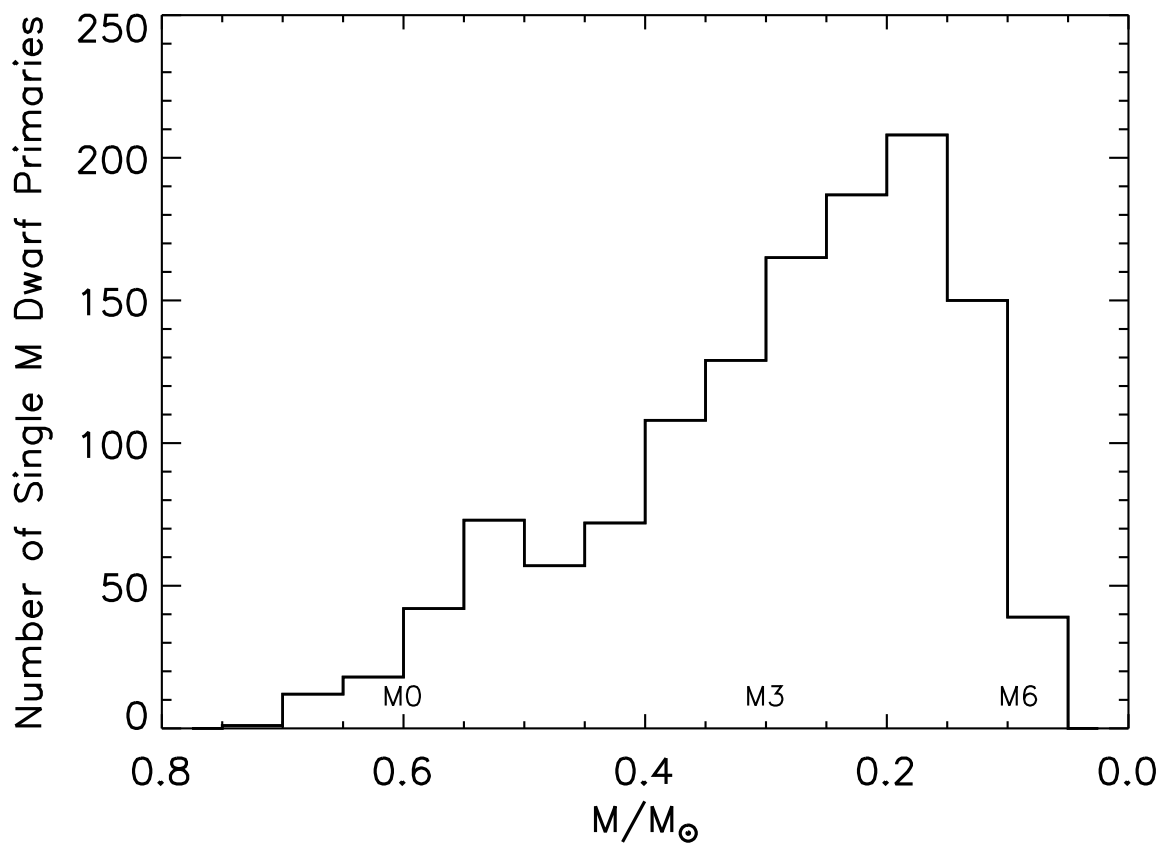}
 
\figcaption[fig11] {The mass function for southern single M dwarf
  primaries known within 25 pc is shown.  Not included are the 100
  primaries with blended photometry due to the presence of a close
  companion. Those objects with mass estimates greater than 0.6
  M$_{\odot}$ are possibly unknown unresolved close multiples. The
  apparent turnover at $\sim$0.15 M$_{\odot}$ is likely not the final
  answer. \label{fig:winters11}}

\end{figure}


\textheight=9.0in
\clearpage

\centering
\begin{deluxetable}{lcccccccccccccccccccc}
\rotate
\setlength{\tabcolsep}{0.03in}
\label{tab:observed}
\tablewidth{0pt}
\tabletypesize{\tiny}
\tablecaption{Observed Data for Southern Red Dwarf Systems}
\tablehead{\colhead{Name}                &
	   \colhead{\# comp.}            &
	   \colhead{RA}                  &
 	   \colhead{DEC}                 &
	   \colhead{$\mu$}               &
	   \colhead{P.A.}                &
           \colhead{ref}                 &
           \colhead{$\pi$}               &
           \colhead{$\pi$ error}         &
           \colhead{\# $\pi$}            &
           \colhead{ref}                 &
           \colhead{$B_J$}               &
           \colhead{$R_{59F}$}           &
           \colhead{$I_{IVN}$}           & 
           \colhead{$V_{J}$}             &
           \colhead{$R_{KC}$}            &
           \colhead{$I_{KC}$}            & 
           \colhead{\# nts/ref}          &
           \colhead{$J$}                 &
           \colhead{$H$}                 &
           \colhead{$K_s$}            \\

	   \colhead{   }                 &
	   \colhead{   }                 &
           \colhead{   }                 &
           \colhead{   }                 &
           \colhead{($\arcsec$yr$^{-1}$)}&
           \colhead{(deg)}               &
           \colhead{   }                 &
	   \colhead{($\arcsec$)}         &
           \colhead{($\arcsec$)}         &
           \colhead{   }                 &
           \colhead{   }                 &
           \colhead{(mags)}              &
           \colhead{(mags)}              &
           \colhead{(mags)}              &
           \colhead{(mags)}              &
           \colhead{(mags)}              &
	   \colhead{(mags)}              &
           \colhead{}                    &
           \colhead{(mags)}              &
	   \colhead{(mags)}              &
           \colhead{(mags)}              \\

	   \colhead{(1)}                 &
	   \colhead{(2)}                 &
           \colhead{(3)}                 &
           \colhead{(4)}                 &
           \colhead{(5)}                 &
	   \colhead{(6)}                 &
           \colhead{(7)}                 &
           \colhead{(8)}                 &
           \colhead{(9)}                 &
           \colhead{(10)}                &
           \colhead{(11)}                &
           \colhead{(12)}                &
           \colhead{(13)}                &
	   \colhead{(14)}                &
	   \colhead{(15)}                &
	   \colhead{(16)}                &
	   \colhead{(17)}                &
	   \colhead{(18)}                &
	   \colhead{(19)}                &
	   \colhead{(20)}                &
	   \colhead{(21)}                }

\startdata

\object[2MASS J00002867-1245153]{2MA0000-1245 }   &1   &00 00 28.7  &$-$12 45 16  &   0.191 & 228.6 & 41  &   ...  &  ...  &...&  ...           &  ...   & 19.26  & 15.88   & 20.78   & 18.71   & 16.27   &   1/   &  13.20   &  12.45   &  11.97    \\ 
\object{LTT09844     }                            &1   &00 00 46.9  &$-$35 10 06  &   0.403 & 106.3 &  1  &   ...  &  ...  &...&  ...           & 13.48  & 11.30  & 10.21   &  ...    &  ...    &  ...    &  ...   &   9.12   &   8.48   &   8.28    \\ 
\object[GJ1001 A]{GJ1001}                         &3(1)&00 04 36.5  &$-$40 44 03  &   1.636 & 159.7 &  1  & .07987 &.00375 & 2 & 25,63          & 14.21  & 11.83  & 10.06   & 12.83   & 11.62   & 10.08   & 4/25u  &   8.60   &   8.04   &   7.74    \\ 
\object{G158-025     }                            &1   &00 04 40.3  &$-$09 52 42  &   0.267 & 170.6 &  1  &   ...  &  ...  &...&  ...           & 14.47  & 12.52  & 10.92   &  ...    &  ...    &  ...    &  ...   &   9.77   &   9.10   &   8.82    \\ 
\object[SIPS J0004-2058]{SIP0004-2058 }           &1   &00 04 41.5  &$-$20 58 30  &   0.826 & 090.6 &  1  &   ...  &  ...  &...&  ...           & 21.86  & 18.85  & 15.65   & 19.94   & 17.81   & 15.36   &   2/   &  12.40   &  11.83   &  11.40    \\ 
\object{GJ0001       }                            &1   &00 05 24.4  &$-$37 21 27  &   6.106 & 112.5 & 28  & .23032 &.00090 & 2 & 63,64          & 10.10  &  7.91  &  6.28   &  8.54   &  7.57   &  6.41   & /5     &   5.33   &   4.83\tablenotemark{c}   &   4.52    \\ 
\object{LP644-034    }                            &1   &00 05 34.9  &$-$06 07 07  &   0.218 & 111.7 &  1  &   ...  &  ...  &...&  ...           & 13.85  & 11.75  &  9.92   &  ...    &  ...    &  ...    &  ...   &   9.26   &   8.65   &   8.41    \\
\enddata
\end{deluxetable}


\centering
\begin{deluxetable}{lccccccccccccc}
\setlength{\tabcolsep}{0.03in}
\label{tab:derived}
\tablewidth{0pt}
\tabletypesize{\scriptsize}
\tablecaption{Derived Data for Southern Red Dwarf Systems}
\tablehead{\colhead{Name}                &
	   \colhead{RA}                  &
 	   \colhead{DEC}                 &
	   \colhead{d$_{plt}$}            &
           \colhead{$\sigma$$_{tot}$}     &
	   \colhead{\# rel}              &
	   \colhead{d$_{ccd}$}            &
           \colhead{$\sigma$$_{tot}$}     &
	   \colhead{\# rel}              &
           \colhead{d$_{trig}$}           &
           \colhead{$\sigma$$_{trig}$}    &
	   \colhead{\# $\pi$}            &
           \colhead{d$_{best}$}           &
           \colhead{d type}              \\

	   \colhead{   }                 &
	   \colhead{   }                 &
	   \colhead{   }                 &
	   \colhead{(pc)}                &
           \colhead{(pc)}                &
           \colhead{   }                 &
	   \colhead{(pc)}                &
           \colhead{(pc)}                &
           \colhead{    }                &
           \colhead{(pc)}                &
           \colhead{(pc)}                &
           \colhead{    }                &
           \colhead{(pc)}                &
           \colhead{    }                \\

	   \colhead{(1)}                 &
	   \colhead{(2)}                 &
           \colhead{(3)}                 &
           \colhead{(4)}                 &
           \colhead{(5)}                 &
	   \colhead{(6)}                 &
           \colhead{(7)}                 &
           \colhead{(8)}                 &
           \colhead{(9)}                 &
           \colhead{(10)}                &
           \colhead{(11)}                &
           \colhead{(12)}                &
           \colhead{(13)}                &
           \colhead{(14)}                }

\startdata

2MA0000-1245  &00 00 28.7  &$-$12 45 16  &    22.2  &  7.6  &  7 &    23.1  &    4.0   & 11 &     ...  &    ...  &... &  23.1    & ccd \\
LTT09844      &00 00 46.9  &$-$35 10 06  &    23.5  &  7.4  & 11 &     ...  &   ...    &... &     ...  &    ...  &... &  23.5    & plt \\
GJ1001        &00 04 36.5  &$-$40 44 03  &    11.2  &  3.4  & 11 &    12.5  &    1.9   & 12 &    12.5  &    0.6  & 2  &  12.5    & trg \\
G158-025      &00 04 40.3  &$-$09 52 42  &    24.5  &  7.4  & 11 &     ...  &   ...    &... &     ...  &    ...  &... &  24.5    & plt \\
SIP0004-2058  &00 04 41.5  &$-$20 58 30  &    14.9  &  6.8  & 10 &    18.6  &    3.0   & 11 &     ...  &    ...  &... &  18.6    & ccd \\
GJ0001        &00 05 24.4  &$-$37 21 27  &     3.7  &  1.0  & 11 &     5.6  &    0.9   & 11 &     4.3  &    0.0  & 2  &   4.3    & trg \\
LP644-034     &00 05 34.9  &$-$06 07 07  &    23.0  &  6.2  & 11 &     ...  &   ...    &... &     ...  &    ...  &... &  23.0    & plt \\
LP644-039     &00 06 13.1  &$-$02 32 11  &    27.3  &  9.2  & 11 &    21.9  &    3.9   & 12 &     ...  &    ...  &... &  21.9    & ccd \\
\enddata

\tablenotetext{a} {The weighted mean distance includes that of
  both the primary and the secondary components.}

\end{deluxetable}


\voffset0pt{
\centering
\begin{deluxetable}{lccccccccc}
\tabletypesize{\scriptsize}
\tablecaption{Comparison Photometry for Giants}
\setlength{\tabcolsep}{0.03in}
\label{tab:giantsdata}
\tablewidth{0pt}

\tablehead{\colhead{Name}                &
           \colhead{RA}                  &
           \colhead{DEC}                 &
           \colhead{$V_J$}               &
           \colhead{$R_{KC}$}             &
           \colhead{$I_{KC}$}            &
           \colhead{\# nts}              &
           \colhead{$J$}                 &
           \colhead{$H$}                 &
           \colhead{$K_s$}               \\

           \colhead{   }                 &
           \colhead{   }                 &
           \colhead{   }                 &
           \colhead{(mag)}               &
           \colhead{(mag)}               &
           \colhead{(mag)}               &
           \colhead{}                    &
           \colhead{(mag)}               &
           \colhead{(mag)}               &
           \colhead{(mag)}               }

\startdata
\object{TT SCL           }&00 16 54.6 &$-$30 13 52 & 11.81 & 10.89 & 10.18 &  2 & 8.92  & 8.21  & 7.98  \\
\object{AF ERI           }&04 20 19.5 &$-$10 24 26 & 13.18 & 10.99 &  8.75 &  2 & 5.80  & 4.94  & 4.45  \\
\object{HD270965         }&05 00 40.4 &$-$71 57 53 & 11.38 & 10.62 &  9.94 &  2 & 8.91  & 8.17  & 7.98  \\
DEN0515-2200              &05 15 36.6 &$-$22 00 10 & 13.37 & 12.27 & 10.81 &  2 & 9.32  & 8.42  & 8.13  \\
\object{HD269328         }&05 17 07.5 &$-$70 56 38 & 10.41 &  9.70 &  9.04 &  1 & 8.05  & 7.33  & 7.20  \\
\object[2MASS J06292015-2101300]{SCR0629-2101}&06 29 20.2 &$-$21 01 30 & 15.37 & 12.98 & 10.84 &  1 & 7.82  & 6.91  & 6.47  \\
SCR0656-2918              &06 56 15.9 &$-$29 18 03 & 13.89 & 11.72 &  9.45 &  1 & 6.37  & 5.51  & 5.05  \\
SCR0703-3507              &07 03 49.6 &$-$35 07 44 & 15.34 & 12.80 & 10.61 &  1 & 7.35  & 6.37  & 5.95  \\
\enddata

\end{deluxetable}
}


\voffset0pt{
\centering
\begin{deluxetable}{lrrccrcclc}
\tabletypesize{\scriptsize}
\tablecaption{Subdwarf Candidates}
\setlength{\tabcolsep}{0.03in}
\label{tab:subdata}
\tablewidth{0pt}

\tablehead{\colhead{Name}                &
           \colhead{RA}                  &
           \colhead{DEC}                 &
	   \colhead{$\mu$}               &
	   \colhead{PA}                  &
           \colhead{d$_{best}$}           &
           \colhead{type}                &
           \colhead{$V_{tan}$}            &
           \colhead{Spec Typ}            &
           \colhead{ref}                 \\

           \colhead{   }                 &
           \colhead{   }                 &
           \colhead{   }                 &
           \colhead{($\arcsec$yr$^{-1}$)}&
           \colhead{(deg)}               &
           \colhead{(pc)}                &
           \colhead{}                    &
           \colhead{(kms$^{-1}$)}         &
           \colhead{}                    &
           \colhead{}                    }

\startdata
LHS0109            & 00 17 40.0  &$-$10 46 17 & 1.051 & 180.0 & 35.21 & trg & 175   & M1.0VI  &   4  \\
LHS0110            & 00 19 37.0  &$-$28 09 46 & 1.388 & 192.2 & 30.45 & trg & 195   & M4.0JVI:&   5  \\
GJ1062             & 03 38 15.7  &$-$11 29 14 & 2.958 & 152.0 & 16.03 & trg & 230   & M2.5VI  &   1  \\ 
LHS0186            & 04 03 38.4  &$-$05 08 05 & 1.153 & 168.0 & 51.55 & trg & 282   & M2.0VI  &   4  \\
WT0135             & 04 11 27.1  &$-$44 18 10 & 0.692 & 067.1 & 25.61 & trg &  84   & M3.0VI  &   5  \\
LHS0189            & 04 25 38.4  &$-$06 52 37 & 1.237 & 145.1 & 18.44 & trg & 108   & M3.0VIJ &   4  \\
GJ0191             & 05 11 40.6  &$-$45 01 06 & 8.728 & 131.4 &  3.91 & trg & 162   & M1.0VI  &   1  \\
LHS0272            & 09 43 46.2  &$-$17 47 06 & 1.439 & 279.2 & 13.52 & trg &  92   & M3.0VI  &   1  \\
\enddata

\tablecomments{Spectral Type References: (1) \citet{Gizis(1997)}; (2)
  \citet{Hawley(1996)}; (3) \citet{Jao(2005)}; (4) \citet{Jao(2008)};
  (5) \citet{Jao(2011)}; (6) \citet{Reid(2005)}; (7)
  \citet{Schilbach(2009)}; (8) \citet{vanAltena(1995)}.}

\end{deluxetable}
}


\voffset0pt{
\centering
\begin{deluxetable}{cccc}
\tabletypesize{\scriptsize}
\tablecaption{Distance Statistics for Southern Red Dwarf Systems}
\setlength{\tabcolsep}{0.03in}
\label{tab:distdata}
\tablewidth{0pt}

\tablehead{\colhead{Distance}                &
           \colhead{\# Systems}              &
           \colhead{\# Multiples}            &
           \colhead{\# Subdwarfs}            }

\startdata
        d $\leq$ 10.0 pc    &  104  &   24 [22]&   1 \\
\hspace*{\fill}                            \\
10.0 $<$ d $\leq$ 25.0 pc   & 1300  &  140 [96] &   8 \\
\hspace*{\fill}                            \\
25.0 $<$ d $\leq$ 100.0 pc  &  343  &   33 [15]&  18 \\
\hspace*{\fill}                            \\
\hline
\hspace*{\fill}                            \\
TOTAL                       & 1747\tablenotemark{a} &   197 &   27\tablenotemark{b} \\
\enddata

\tablecomments{Numbers in brackets indicate systems with unresolved
  photometry, some of which were used to calculate distance estimates
  that are likely underestimates.}

  \tablenotetext{a} {CD-32 16735 is beyond 100 pc and is not included
    in this total.}  \tablenotetext{b} {Two of the subdwarf systems
    are binaries, and thus are also included in the ``Multiples'' counts
    column.}

\end{deluxetable}
}

\end{document}